\documentstyle[epsf,mnras_cite]{mn}
\def\spose#1{\hbox to 0pt{#1\hss}}
\def\simlt{\mathrel{\spose{\lower 3pt\hbox{$\mathchar"218$}}
     \raise 2.0pt\hbox{$\mathchar"13C$}}}
\def\simgt{\mathrel{\spose{\lower 3pt\hbox{$\mathchar"218$}}
     \raise 2.0pt\hbox{$\mathchar"13E$}}}
\def\et{{\it et~al. }}
\def\hmpc{\;h^{-1}{\rm Mpc}}

\def\kms{{\rm \,kms}^{-1}}
\def\kmsmpc{\kms\;{\rm Mpc}^{-1}}

\def\etal{{\it et~al. }}

\def\hmpcrev{\;h {\rm Mpc}^{-1}}

\def\be{\begin{equation}}
\def\ee{\end{equation}}
\def\sumlmn{\sum_{\ell m n}}

\def\r{\mbox{\boldmath $r$}}
\def\s{\mbox{\boldmath $s$}}
\def\v{\mbox{\boldmath $v$}}
\def\vb{\mbox{\boldmath $v$}}
\def\vlg{{\vb}_{\rm LG}}
\def\rhat{\hat{\r}}
\def\W{\mbox{\boldmath $W$}}
\def\bPhi{\mbox{\boldmath $\Phi$}}
\def\bV{\mbox{\boldmath $V$}}

\def\bdelta{\mbox{\boldmath $\delta$}}
\def\Smat{\mbox{\boldmath $S$}}
\def\N{\mbox{\boldmath $N$}}
\def\D{\mbox{\boldmath $D$}}
\def\A{\mbox{\boldmath $A$}}
\def\k{\mbox{\boldmath $k$}}
\def\C{\mbox{\boldmath $C$}}
\def\R{\mbox{\boldmath $R$}}
\def\bP{\mbox{\boldmath $P$}}
\def\Ylm{Y_{\ell m}(\theta,\phi)}
\def\lgl{\langle}
\def\rgl{\rangle}

\newcommand{\ba}{\begin{eqnarray}}
\newcommand{\ea}{\end{eqnarray}}

\newcommand{\nn}{\nonumber \\}
\newcommand{\Sym}{{\rm Sym}}
\newcommand{\Asym}{{\rm Asym}}
\newcommand{\Real}{\Re \!e}
\newcommand{\Imag}{\Im \!m}
\begin{document}

\title[
Spherical Harmonic Analysis of the PSCz Galaxy Catalogue 
]
{
Spherical Harmonic Analysis  of the  PSCz  Galaxy  Catalogue: 
Redshift distortions and the real--space power spectrum
}
\author[H. Tadros {\it et al.}]{H. Tadros,$^{1,2}$
W. E. Ballinger,$^{3,2}$ A. N. Taylor,$^3$, A. F.
Heavens,$^3$ G. Efstathiou$^4$,\\
\vspace{-1mm}\\ 
{\LARGE  W. Saunders$^3$, C. S. Frenk$^{5}$, O. Keeble$^{6}$, 
R. McMahon$^4$, S. J. Maddox$^4$,
},\\
\vspace{-1mm}\\ 
{\LARGE
S. Oliver$^6$, M. Rowan-Robinson$^6$, W. J. Sutherland$^2$ and
S. D. M. White$^7$}\\ 
$^1$Astronomy Centre, University of Sussex,
Falmer, Brighton UK.\\ 
$^2$Department of Physics, University of Oxford, Keble Road, Oxford, UK.\\ 
$^3$Institute for Astronomy, University of Edinburgh, Blackford Hill, 
Edinburgh.\\
$^4$Institute of Astronomy, Cambridge, UK. \\
$^5$Department of Physics, University of Durham, UK. \\
$^6$Imperial College, University of London, UK.\\
$^7$MPI-Astrophysik, Garching, Germany.\\
}

\maketitle

\begin{abstract}
We apply the formalism of spherical harmonic decomposition to the
galaxy density field of the IRAS PSCz redshift survey. The PSCz redshift 
survey has almost all--sky coverage and includes IRAS galaxies to a flux 
limit of $0.6$ Jy.  Using maximum likelihood methods to examine (to first 
order) the distortion of the 
galaxy pattern due to redshift coordinates, we have measured the parameter
$\beta\equiv \Omega^{0.6}/b$. We also
simultaneously measure (a) the undistorted amplitude of
perturbations in the galaxy distribution when a parameterised power
spectrum is assumed, or (b) the shape and amplitude of the real--space
power spectrum if the band--power in a set of passbands is measured in
a step--wise fashion.  These methods are extensively tested on a
series of CDM, $\Lambda$CDM and MDM simulations and are found to be
unbiased.

We obtain consistent results for the subset of the PSCz catalogue with
flux above 0.75 Jy, but inclusion of galaxies to the formal flux limit
of the catalogue gives variations which are larger than our internal
errors.  For the 0.75 Jy catalogue we find, in the case of a
parameterised power spectrum, $\beta= 0.58\pm 0.26$ and the amplitude
of the real space power measured at wavenumber $k=0.1h{\rm Mpc}^{-1}$
is $\Delta_{0.1}=0.42 \pm 0.03$. Freeing the shape of the power
spectrum we find that $\beta=0.47\pm 0.16$ (conditional error), and
$\Delta_{0.1}=0.47 \pm 0.03$. The shape of the real--space power
spectrum is consistent with a $\Gamma=0.2$ CDM--like shape parameter,
but does not strongly rule out a number of other models. Finally by
combining our estimate of the amplitude of galaxy clustering and the
distortion parameter we find the amplitude of mass fluctuations on a
scale $k=0.1 \hmpc$ is $\Delta_\rho = 0.24 \Omega_0^{-0.6}$, with an
uncertainty of $50\%$.
\end{abstract}

 
\section{Introduction}
\label{ssintro}

The pattern of galaxy clustering in redshift surveys is distorted because
of peculiar (non-Hubble) velocities which contribute to the line of sight 
velocity measured by the galaxy redshift. This
distortion makes the interpretation of the clustering more
difficult, and must be corrected for when measuring  statistical
properties of the pattern, such as the 2-point correlation function
or power spectrum. However the distortion itself contains important
information about the motion of matter in the universe. Because the
effect is only radial, the measured clustering pattern becomes anisotropic. 
The degree
of anisotropy is determined by the distortion parameter\footnote{
A distinction must be made between the distortion parameters for
different galaxy types, since their relative biases are known to differ. 
Since we are concerned with IRAS selected galaxies in this paper, we
 shall reserve $\beta$ and $b$ to refer to IRAS galaxies.}
$\beta\equiv \Omega^{0.6}/b$, where $\Omega$ is the cosmological 
mass--density parameter and $b^2 \equiv P(k)_{\rm gal}/P(k)_{\rm mass}$ is the 
bias parameter, relating clustering in mass to clustering in 
galaxies.

By measuring the degree of distortion one can hope to constrain $\beta$,
and hence the density parameter and bias.
In addition,  measurement of the real--space (i.e. undistorted) power spectrum can put 
strong constraints on theories of structure formation. A major aim
of large--scale redshift surveys is to constrain the shape and
position of the break in the power spectrum as it turns over
into the primordial spectrum. This would demonstrate a strong link
between the clustering responsible for the microwave background
fluctuations and that found in redshift surveys, as well as
putting strong constraints on the slope of the connecting spectrum.
This in turn provides constraints on inflationary models for the
generation of structure.

A comprehensive overview of the literature to date concerning the analysis of redshift space distortions is given  by \scite{ham97}. Here we shall outline some of the main results
leading up to this analysis.

\scite{Jackson72} first noticed the distortion when viewing a cluster
of galaxies in redshift--space, where peculiar velocities are
generated by virial motion. \scite{Kai87} derived the main features of
the redshift distortion in the linear regime, simplifying the analysis
by assuming plane density waves and a distant observer
approximation. The major result was that power is boosted in redshift
space by a factor $(1+\beta \mu^2_k)^2$, where $\mu_k$ is the cosine
of the angle between wavevector and line of sight. \scite{HAM93}
suggested expanding the distorted power in Legendre polynomials, since
the main distortion is quadrupole in nature. Taking the ratio of
quadrupole to monopole power yields a function in terms of $\beta$
only, and one could hope to measure the distortion independently of
power.  Applying this to the IRAS 2Jy survey \scite{HAM93} found
$\beta=0.69 \pm0.26$.  Hamilton further applied this method to a
merged QDOT + 1.2 Jy redshift survey and found $\beta=0.69
\pm0.20$. However Hamilton also suggested that the merged survey was
systematically biased due to a difference in clustering between the
near and far regions of the QDOT sample.

\scite{CFW94} and \scite{CFW95} tested the quadrupole to 
monopole estimator on simulations and applied it to the IRAS 1.2Jy and QDOT 
surveys and found $\beta=0.52\pm0.15$. \scite{PD94}
used a radially projected Gaussian smoothing function to model the
effects of nonlinearity in the small scale velocity field. They found
that $\beta=1.0\pm0.2$ for IRAS galaxies, using the APM survey to constrain
the real-space power spectrum (\pcite{BEI}), and allowing a relative 
bias between the galaxy samples.

\scite{th96} and \scite{fn96}  suggested that
the main contribution to nonlinearity in redshift space was not from
an incoherent velocity dispersion, but rather from coherent infall
in the translinear regime. Using the Zel'dovich approximation 
they calculated the scale dependence of the quadrupole to monopole 
ratio, finding that the amplitude was still only a function of
$\beta$. \scite{th96} also found a dependence on the initial
power spectrum, and the scaling behaviour of the wavenumber at which the 
quadrupole is zero. They applied these results to the merged QDOT + 1.2Jy 
surveys and found $\beta=0.6\pm0.2$, assuming a local spectral index
of $n=-1.5$, while \scite{fn96} found $\beta=0.6\pm0.2$
from the 1.2Jy survey alone.

\scite{FSL94} and \scite{HT}
both dropped the plane parallel approximation and used a spherical
harmonic decomposition to match the spherical nature of the IRAS
redshift surveys. \scite{FSL94} used a Gaussian
window function in the radial direction and, applying a maximum likelihood
formalism, found $\beta =0.96\pm0.19$ for the 
1.2Jy survey. \scite{HT} used spherical Bessel functions
to decompose the density field radially, since these combine the property 
of being orthonormal and eigenfunctions of the Laplacian, with the
advantage that each mode only picks up power from a narrow range of 
wavenumbers. In addition 
they applied a radial smoothing to correct for the effects of nonlinearity.
Applying a maximum likelihood analysis they measured simultaneously
$\beta$ and the amplitude of a fixed power spectrum (characterised by the 
fractional r.m.s. in counts in 8$h^{-1}$Mpc spheres), 
finding $\beta=1.1\pm0.5$
and $\sigma_8=0.68\pm0.05$, where we quote marginal errors.  
\scite{Ballinger} extended 
the analysis by dropping the fixed power spectrum and measuring 
the real--space band--power in a series of passbands in a step-wise fashion.
With a free power spectrum, they found $\beta=1.04$ with a conditional
error of $0.3$.

In this paper we shall apply the methods
of \scite{HT} (hereafter HT) and \scite{Ballinger} (hereafter BHT)
to measure the degree of distortion and the amplitude and shape of
the undistorted power spectrum of the IRAS Point Source 
Catalogue redshift survey (PSCz; \pcite{SPSCZ98}).

The contents of the paper are as follows. In Section 2 we discuss the
methods used, including some technical improvements to the original
analysis.  In Section 3 we extensively test the methods on a suite of
simulated PSCz redshift surveys. In Section 4 we give a brief outline
of the PSCz catalogue itself and highlight the checks made on possible
sources of contamination to the present analysis. In Section 5 we
present the results of the analysis for the redshift distortion and
amplitude of the real--space power spectrum, and the shape of the
real--space power spectrum measured in a series of passbands. In
Section 6 we discuss the implications of these results and present our
conclusions.  There are also two technical appendices outlining the
spherical harmonic formalism, deriving the main equations used in the
analysis and detailing the construction of the mixing matrices. We
also explain, in more detail, the technical improvements made to the
analysis in the present work. We begin by discussing the methods to be
used.

\section{Methods}
In this paper we present two complementary analyses of the redshift distortions
in the PSCz survey. The first analysis, described in detail in Section
2.1, is a two parameter maximum likelihood analysis of the data to obtain
the value of $\beta = \Omega^{0.6}/b$ and the amplitude of the real
space power spectrum at a wavenumber $k = 0.1 \hmpcrev$\footnote{
Throughout this paper we shall use $h=H_0/100\kmsmpc$, where $H_0$ is the 
Hubble parameter.}
 (assuming 
a parametrised form for the shape of the power spectrum). The details
of this analysis follow HT with some technical improvements to the
method concerning how the sky mask and Local Group motion are dealt with. 
These modifications
are explained in Section 2.1 and in Appendix A. The second analysis, 
described in Section 2.2 follows BHT and  is an extension of
that described above. Here we perform a likelihood fit for $\beta$ and 
the full real space power spectrum, sampled in a series of passbands. This
has the advantage of relaxing the assumption of a fixed spectral shape.
We begin by discussing the galaxy density fluctuations in redshift--space.

\subsection{Density fluctuations in redshift--space}

Following HT, the density field of the galaxy distribution $\rho(\s)$
is expanded in terms of spherical harmonics, $Y_{\ell m}$, and
a discrete set of spherical Bessel functions, $j_{\ell}$,
\begin{equation}
	\hat{\rho}_{\ell mn} = c_{\ell n} \int \! d^3s \, 
	\rho(\s) w(s)j_{\ell}\left(k_{\ell n}
	s\right)
	Y^{*}_{\ell m}\left(\theta,\phi\right),
\label{transequ}
\end{equation}
where $w(s)$ is an adjustable weighting function and $\s$ is the
redshift-space position variable. The inverse transform is
\be
	\rho(\s) = \sumlmn c_{\ell n} \rho_{\ell m n} j_\ell(k_{\ell n} s) 
			Y_{\ell m}(\theta, \phi),
\label{invtransequ}
\ee
where the $c_{\ell n}$ are normalization constants and $k_{ln}$ are
discrete wavenumbers (see Appendix C of HT and Appendix A in this work
for a definition).

The domain of the integral in equation (\ref{transequ}) can be
transformed from redshift space $\s$ to real space $\r$ by continuity, $d^3s\,
\rho(\s) = d^3r\, \rho(\r)$, leaving only the argument of the
spherical Bessel function as a function of the redshift-space
distance $s$. Noting that redshift distortions only affect radial
coordinates, the transformation can be written 

\be s(\r) = r +
\frac{u(\r)}{H_{0}}, 
\ee 
where $u=\rhat. (\vb(\r)-{\vb}_{\rm LG})$
is the radial component of the peculiar velocity field in the Local
Group frame and $H_{0}$ is the Hubble parameter.  Note that the
expansion of the density field is done in this frame, since we need to
assume that $|u|\ll H_0 r$, which will only be the case in the near-field
limit if we choose a frame in which $\vb - {\vb}_{\rm frame}
\rightarrow {\bf 0}$ as $r \rightarrow 0$.  In linear theory the
velocity field is related to perturbations in the mass density by the
continuity equation 
\be \nabla. \v = - H_0 \Omega^{0.6} \delta_\rho, 
\ee
where $\Omega^{0.6} \approx d \ln \delta_\rho /d\ln a$ is the linear
growth index and $\delta_\rho$ is the fractional matter overdensity.
Expanding the continuity equation in spherical harmonics we find that
the radial velocity field can be expressed as 
\be u(\r) = \Omega^{0.6}
\sumlmn c_{\ell n} \delta_{\ell m n } k^{-1}_{\ell n} j'_\ell(k_{\ell
n}r) \Ylm, 
\ee 
where $j'_\ell(z) \equiv d j_\ell(z) /d z$.

Expanding equation (\ref{transequ}) to first order in $u=r-s$ we find
\be
	\hat{\rho}_{\ell mn} = \left(\rho_{0}\right)_{\ell mn} +
	\sum_{\ell' \!m'\! n' }
	W_{\ell \ell'}^{mm'}\left(\Phi^{nn'}_{\ell \ell'} +
	\beta V^{nn'}_{\ell \ell'}\right)
	\delta_{\ell'\!m'\!n'},
\ee
where the mean coefficients $\left(\rho_{0}\right)_{\ell mn}$ are given 
in Appendix A, and include a term for the Local Group velocity, although 
the effect of this for $\ell\ne 1$ is very small.  
The transition matrices $\W$, $\bPhi$, and $\bV$ describe the effects 
of the sky mask, the radial selection function and the first order redshift 
space distortion, respectively. The transition matrices are  derived and 
defined in Appendix A.

In HT it was assumed that the sky mask was azimuthally symmetric. This 
simplified the analysis since the mask mixing matrix, $\W$, was then 
real. Given the higher surface density of galaxies in the PSCz we drop this 
assumption and use the exact PSCz mask (see Figure~\ref{psczsky}). Also, 
we do not apply an artificial galactic cut, as was done conservatively 
for the 1.2Jy survey (HT). The mask used for this analysis is 
described briefly in Section 4 and in Appendix B we describe in detail the 
methods used for mask
generation and application.

Throughout this paper we have also applied a radial correction for the 
effects of small--scale nonlinear peculiar velocity fields (see HT, Section 4.4
and Appendix B, therein, and Appendix A in this work). This assumes that the 
effects of the `fingers of God'
due to virialised clusters and redshift measurement errors can be
modelled by adding an incoherent random displacement to each galaxy
radial distance
\be
	s' = s + \varepsilon(\r),
\ee
where $\lgl \varepsilon \rgl = 0$,
$\lgl \varepsilon(\r) \varepsilon(\r') \rgl = (\sigma^2_v/3H_{0}^{2})
\delta_D(\r-\r')$ and $\sigma_{v}$ is the three-dimensional velocity dispersion.
If the distribution of random displacement is assumed to be drawn from
a Maxwellian distribution we should ensemble average 
over this distribution when 
forming the correlators of the density field, $\lgl \delta \delta' \rgl$.
But since the random displacements are incoherent we need only 
calculate the effects on the mean value of modes, since 
$\lgl X Y \rgl = \lgl X \rgl \lgl Y \rgl$ if $X$ and $Y$ are uncorrelated.
The averaging effect on a given mode can be expressed as a convolution of
the density field by a scattering matrix, $\Smat$ (see Appendix A for a definition).

The statistical properties of the redshift space density field can be 
encapsulated by the mode--mode correlation function 
\be
	\lgl \D \D^\dag \rgl = 
	\frac{1}{2} \sum \Smat  \W (\Phi + \beta \bV) \bP  
	(\Phi + \beta \bV)^t \W^\dag \Smat^t
	+ \N \nn
\label{modemode}
\ee
where the observable data vector, $\D$, is given by 
$D_{\ell mn} \equiv [\hat\rho_{\ell mn}-(\rho_0)_{\ell mn}]/\bar\rho$ and
$\bar\rho$ is the mean number density, and $\N$ is the shot--noise
contribution to the covariance matrix.  We have assumed that the
underlying Fourier modes of the density field are statistically
isotropic and homogeneous:
\be
	\lgl \delta(\k) \delta^*(\k') \rgl = (2 \pi)^3 P(k) \delta_D(\k-\k'),
\ee
where $\delta_D$ is the Dirac delta function.  Since the shot--noise
term contains a part linear in the matrix $\W$, the mode--mode
correlator is complex for a non-azimuthally-symmetric mask.  This can
be easily overcome by decomposing the transformed density field into
real and imaginary parts:
\be
	\D = \Real \D + i \Imag \D.
\ee
This is convenient since our hypothesis that $\D$ is a Gaussian random
variable requires that the terms $\Real \D$ and $\Imag \D$ are also Gaussian
random variables. In the limit of full sky coverage these terms would also be
uncorrelated, but the presence of a non-azimuthally symmetric mask 
mixes real and imaginary parts. Even so the likelihood function for 
the observed redshift density field can be written in Gaussian form:
\begin{equation}
	{\cal {L}}[D|\beta,P(k)] = \frac{1}{(2 \pi)^{N/2} |{\C}|^{1/2}}
	{\rm exp}\left[ - \frac{1}{2} \D^t \C^{-1} \D\right],
\end{equation}
where $\D=(\Real \D, \Imag \D)$ is a data vector of real and 
imaginary parts of the density modes, with dimension its
 being set by the upper limit to the wavenumber.  The 
data covariance matrix is
\be
		\C =    
        \left(
        \begin{array}{cc}      \C_{\Real \Real} & \C_{\Real \Imag} \nn
                               \C_{\Imag \Real} & \C_{\Imag \Imag}  
        \end{array}
         \right)
\label{cov}
\ee
where $\C_{\Real \Real} = \lgl \Real \D \, \Real \D^t \rgl$, 
$\C_{\Real \Imag} = \lgl \Real \D \, \Imag \D^t \rgl$ and so on.
From the symmetry relations given in Appendix B we see that this is 
a real, symmetric matrix. 

Another refinement to the analysis of HT is the inclusion of $m=0$ modes.
Since these are real modes, which have twice the power as the $m\neq0$
modes, we treat these separately. 

Finally, we have also included a correction for the effect of the Local
Group motion. In HT and BHT, we simply removed the dipole contribution
to the harmonics. While this is exact for full sky coverage, for
incomplete sky coverage there is a mixing of modes and the dipole 
leaks into higher modes, although at a level which was negligible for 
the PSCz mask used. Here we account for that leakage. Again 
technical details can be found in Appendix A.

The free parameters for the likelihood are then 
the redshift distortion parameter $\beta$ and the real--space power
spectrum $P(k)$. In this paper we shall consider simultaneous
measurement of 
\begin{description}
\item[(a)] $\beta$ and the normalisation of $P(k)$ for 
a fixed spectral shape, 
\item[(b)] $\beta$ and the shape and amplitude
of $P(k)$. 
\end{description}
The first method maximises the likelihood over a 2-dimensional
parameter space, while the latter requires a maximisation over a 7-dimensional
parameter space -- one dimension for $\beta$ and 6 dimensions for the 
step-wise power spectrum sampled uniformly in logarithmic pass--bands. In 
Section 2.2 we shall discuss the step--wise maximum likelihood method
for measuring the real power spectrum. Before doing so there
are two issues we shall deal with here -- that of the correct weighting 
function
to use in the transformation equation (\ref{transequ}) and the choice of 
how to 
measure the normalisation of the power spectrum.

The minimum variance weighting scheme to use in the transformation equation 
(\ref{transequ})
has been discussed by \scite{FKP94}, HT, and \scite{ham97}.
All three have found that for an undistorted, flux limited
redshift survey the optimal weighting function for measuring power spectra is
\be
	w(r) = \frac{1}{1+\phi(r) P_w(k)},
\label{weight}
\ee
where $\phi(r)$ is the survey selection function in real space and $P_w(k)$ is
the expected power spectrum.   As can be seen one has to make an initial
guess for the power spectrum to insert into the weighting function.  This
is not important: a poor choice would lead to larger than necessary error bars,
but should not bias the solution.  In any case, if one was concerned, one could
iterate. We have chosen to use a $\Gamma=0.2$ power spectrum for the analysis 
of the PSCz, where $\Gamma$ is the CDM-like spectral shape parameter. In 
the tests on
mock PSCz catalogues (Section 3) we used the $P_w(k)$ appropriate to the 
simulations. In both cases we have used the full $k$-dependence in the
weighting scheme (cf HT, who used a $k$-independent P(k) in the analysis
of the 1.2Jy survey).  Note that this weighting function is strictly only 
optimal for estimating power spectra, and when $\beta=0$.  HT derived
an approximate weighting scheme optimised for measuring $\beta$, with
the main difference being to correct for the misplacement of weights in 
redshift space and to boost the power by a factor $(1+\beta)$.  
Since these seem to have little effect on the maximum likelihood solution, 
we used the weighting function optimised for power.

In the two-parameter fits, we have chosen to express the power normalisation 
in terms of $\Delta_{0.1}$, where 
\be	
	\Delta^2_k\equiv P(k)k^3/(2\pi^2)
\ee
is the contribution to the overdensity variance per ln($k$).  This choice is
motivated by the fact that we choose an upper wavenumber limit 
$k\simeq 0.1h {\rm Mpc}^{-1}$ to avoid nonlinear scales, and most of the 
modes are concentrated near the upper wavenumber limit.  Hence it is the
power at this wavenumber which is most strongly constrained and 
is the most insensitive to the shape of the power spectrum assumed.  
Since it is common to normalise spectra by the variance in spheres of 
radius $8 \hmpc$, 
\be
	\sigma^2_8 = \int \frac{dk}{k} \Delta^2(k) 
	\left[\frac{3 j_1(k r_8)}{ kr_8} \right]^2,
\ee
we have fitted the 
relationship between $\sigma_8$ and $\Delta_{0.1}$ as a function of
the CDM shape parameter, $\Gamma$. A fit accurate to a few percent is given 
by
\be
	\sigma_8^2 = 
	0.616 \left[1 - 0.84 \left(1 - \frac{\Gamma}{0.2}\right) \right]
	 \left( \frac{\Delta_{0.1}}{0.5}\right)^2.
\label{s8}
\ee

A further consideration is the validity of the linear bias model used 
to relate the galaxy distribution to the peculiar velocity field, which 
appears in $\beta$. While a wavenumber of $k=0.1 h{\rm Mpc}^{-1}$ corresponds
to a scale of about $60 \hmpc$, and one would hope biasing could be 
described by 
a local and linear model, we have no concrete guarantee that biasing 
is not still scale--dependent or non-local, as in the case of 
cooperative biasing (\pcite{BCFW93}). However
realistic models of biasing mechanisms do seem to give some support 
to local biasing that is approximately linear on these scales (\pcite{Kauff97},
\pcite{Mann98}, \pcite{HMV98}).

Having described the main formalism for the measurement of distortions and
real space power in this section, we now turn to the methods
used to measure the shape and amplitude of the power spectrum in a stepwise
fashion.

\subsection{Stepwise maximum likelihood}

This method fits the shape of a function in an essentially 
model-independent way -- see \scite{EEP88}. The function is divided 
into bins, and the amplitude of these bins is allowed to vary.


We exploit the fact that we have a good estimate for the power spectrum from 
previous studies, $P_{0}(k)$, so we write 
\be
P(k) = P_0 (k) \times 10^{f_i},
\label{equPk}
\ee
where the original power spectrum (CDM with $\Gamma=0.2$ and 
$\Delta_{0.1}^2=0.18$) is multiplied by 
the free function $10^{f_i}$.
The subscript $i = 1,2,3,\ldots$ labels each wavenumber bin corresponding 
to the range $k_i^{\rm min} < k < k_i^{\rm max}$.
 The parameters are introduced in this exponential form to ensure that the 
measured power spectrum is positive.

The parameters
for the maximum likelihood fit are the distortion parameter $\beta$ and
the coefficients $f_i$; $i=1,\ldots,n$.  This construction essentially
amounts to specifying the power spectrum at a series of wavenumbers,
but the form chosen allows us to start the maximum likelihood search
conveniently at $f_i=0$, since we have a fairly good idea of what to
expect.  Note that, although the form suggests that the break scale is
fixed,  adjustment of the $f_i$ can alter this in an essentially
arbitrary way, which is the main reason for parametrising the power spectrum 
in bins.   
In HT, and in \scite{FSL94}, the power spectrum was parametrised 
to have a specific shape, whereas here it has a free form 
(apart from bin discreteness).

We begin by expanding the covariance matrix of equation (\ref{cov}) in 
powers of $\beta$;
\be
	\C = \sum_{n=0}^2 \beta^n \A_n + \N
\ee
The covariance matrix can now be constructed with little 
computational cost as each sub-matrix can be written as a linear 
combination of contributions from each power-spectrum bin;
\ba
\A_0 &=& 10^{f_1}\A_{01} + 10^{f_2}\A_{02} + \ldots, \nn
\A_1 &=& 10^{f_1}\A_{11} + 10^{f_2}\A_{12} + \ldots, \nn
\A_2 &=& 10^{f_1}\A_{21} + 10^{f_2}\A_{22} + \ldots, 
\ea
where $\A_{01}$ is the contribution to $\A_0$ from the first bin, and so on. 
The matrices $\A_{0i}$, $\A_{1i}$ and $\A_{2i}$ need only be constructed 
once and then the full A-matrices can be created each time from a simple 
linear sum. This requires far less computer time than fitting 
for a standard power spectrum parameter such as $\Gamma$. That would require 
the A-matrices to be constructed afresh from window function convolutions 
each time a parameter changed.

The maximizing procedure is straightforward, 
the $(n+1)$D
parameter space is searched until a maximum of the likelihood function
${\cal L}[D|\beta,f_i; i=1,\ldots,n]$ is found.   
This generalizes the method of HT, where, in effect, the $f_i$ were all 
constrained to move up or down together.

If $n$ is high, exploring the likelihood space with a grid can become
computationally expensive. However, the maximum likelihood solution is
easily found, since we have a good starting guess for the parameters,
and can use a standard library routine to maximise the likelihood
function with respect to the parameters.  Ideally, we would like to
construct complete hypersurfaces of constant likelihood, but this is
impractical because of the dimensionality of the problem.  Errors can
be obtained by searching along the parameter axes until the likelihood
drops by $\sqrt{{\rm e}}$, but note that this ignores the correlations
between parameters. This means that the errors are {\em conditional}:
for each parameter they are calculated assuming that the other
parameter values are known exactly -- see Press \etal\
(1992; \S15) for a discussion.  Note in particular that there is a
strong covariance between the amplitude of the power spectrum and
$\beta$, and these errors do not properly reflect this (see HT,
\pcite{FSL94}).

By parametrising in bins, the fitted power
spectrum is discontinuous.  
In principle there might be a case for adjusting the shape 
of the power spectrum so that it passes continuously through the 
maximum likelihood points;  
in the event there is little to be achieved by this, since the error bars 
are large enough so that a smooth curve is a perfectly acceptable fit.

\section{Tests on Simulated Surveys}

The methods described in Sections 2.1 and 2.2 have been applied to mock
PSCz redshift surveys drawn from large N-body simulations. Before
presenting the results from the PSCz redshift survey, we describe the
result of applying these methods to the mock PSCz data sets.

\subsection{Numerical simulations}
The dissipationless N-body simulations that we have used to create the mock 
data-sets consist of 3 ensembles 
of 9 realizations each. The
first two ensembles are of CDM-like universes and contain 
$160^{3}$ particles within a periodic
computational box of length $\ell_b = 600 \hmpc$.  These simulations
are similar to the ones used and discussed in detail by \scite{CE94a}
but employ more particles within a larger computation box (the
simulations of Croft and Efstathiou use $100^{3}$ particles within a
box of size $\ell_b = 300 \hmpc$).  The simulations were run with the
particle-particle-particle-mesh (P$^{3}$M) code described by
\scite{EDFW85} and model the gravitational clustering of dark matter in a 
CDM dominated
universe with scale invariant initial density fluctuations.  These two
ensembles are as follows: the standard CDM model (\pcite{DEFW85}),
{\it i.e.} a spatially flat universe with $\Omega_{0}=1$ and $h=0.5$
(the SCDM ensemble); a spatially flat low density CDM universe with
$\Omega_{0}=0.2$, $h=1.0$, and a cosmological constant contribution
$\Omega_{\Lambda}= \Lambda /(3H_{0}^{2})
=\left(1-\Omega_{0}\right)=0.8$ (the $\Lambda$CDM ensemble).  The initial
power spectra of the models are generated from the fitting function
\begin{equation}
	P\left(k\right)\propto
	\frac{k}{\left[1+\left[ak+ \left(bk\right)^{\frac{3}{2}}+
	\left(ck\right)^{2} \right]^{\nu}\right]^{\frac{2}{\nu}}},
\label{CDM}
\end{equation}
where $\nu=1.13$,
$a= (6.4/\Gamma) \hmpc$, $b=(3.0/\Gamma)\hmpc$ and
$c=(1.7/\Gamma) \hmpc$. Equation (18) is a good
approximation to the linear power spectrum of scale-invariant CDM
models with low baryon density, $\Omega_{b}\ll \Omega_{0}$
(\pcite{BE84}). The parameter $\Gamma$ in equation
(18) is equal to $0.5$ for the
SCDM ensemble and $\Gamma = 0.2$ for the $\Lambda$CDM ensemble.
For pure CDM-type models the shape parameter is given by $\Gamma=\Omega h$.
But if the model contains baryonic matter the shape parameter can be 
rescaled so $\Gamma=\Omega h e^{-\Omega_B/2}$, where $\Omega_B$ is the
baryon density parameter.
\cite{Peacock92}.

The third ensemble of $9$ simulations is a spatially flat mixed dark
matter model (the MDM ensemble) in which CDM contributes $\Omega_{\rm CDM}=
0.6$, baryons
contribute $\Omega_b = 0.1$ and massive neutrinos contribute
$\Omega_\nu = 0.3$. This set of
simulations was run in a box of length $300 \hmpc$ with $100^{3}$
particles. The initial conditions for these simulations are generated
from the power spectrum given in equation (1) of \scite{KHPR93}. In
the MDM models, the thermal motions of the neutrinos are ignored and
so they follow the evolution of a collisionless cold component with
$\Omega_0 = 1$. The MDM simulations are exactly those used by
\scite{CE94b}.

In the analysis presented below, we use data and simulated data out to
a comoving distance of at least $240 \hmpc$, this means that the MDM
models are not large enough to properly simulate the survey
without a significant amount of wrapping of the simulation box. The
larger CDM-like models however are superior in this respect and no
wrapping is required to simulate the data.
 
The final output times of the models are chosen to approximately match
the microwave background anisotropies measured in the first year COBE
maps (\pcite{COBE}) ignoring any contribution from gravitational
waves; 
\be
	\Delta(k)=180 k^2 T(k).
\ee
 Thus the {\it rms} mass fluctuations 
in spheres of radius $8
\hmpc$ are $\sigma_8 = 1$ for the CDM-like models and $\sigma_8 =
0.67$ for the MDM model. The amplitude of the temperature anisotropies
measured from the four year COBE maps (\pcite{Wright96}) is slightly
higher than from the first year results. However, for the purposes of
testing the techniques described above, this discrepancy is
unimportant.

\begin{figure*}
\centering
\begin{picture}(400,400)
\includegraphics{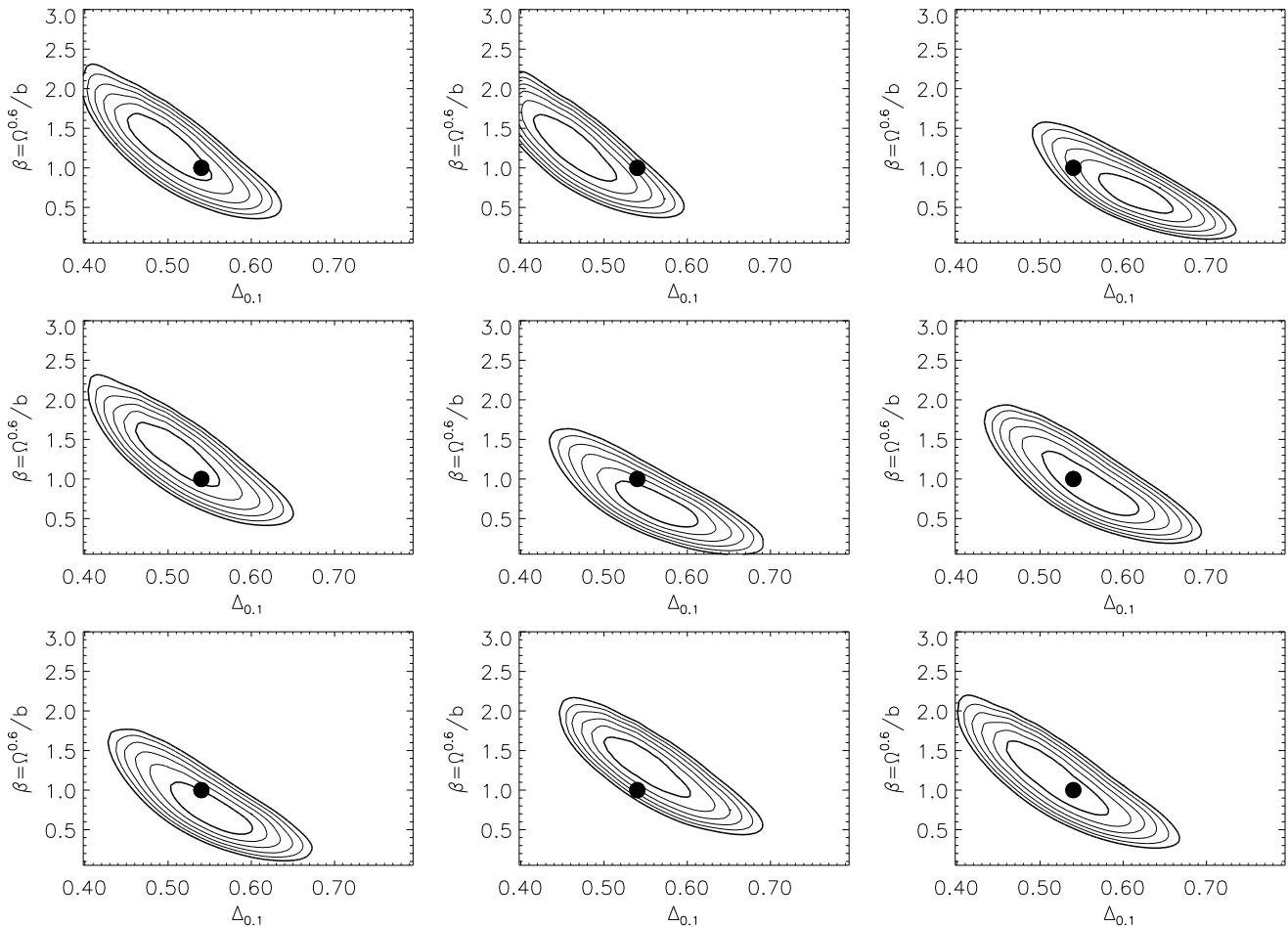}
\end{picture}
\caption[]{\label{SCDM} Figure shows contours of likelihood in the
$\beta$ -- $\Delta_{0.1}$ plane for 9 realizations of the PSCz
redshift survey drawn from SCDM N-body simulations.  Points show the
true parameter values.  The contours are
plotted at intervals
of $\delta \ln {\cal{L}} = 0.5$.  The joint distribution has ($\beta$, 
$\Delta_{0.1}$) within 2.3 contours with 68\% probability, and within 
6.17 contours with 95\% probability.  The 1-$\sigma$ error bars on the 
parameters individually are obtained by projecting the inner contour onto
the axes \cite{Press92}.}
\end{figure*}

\subsection{Construction of mock PSCz surveys}
We draw mock PSCz redshift surveys from the N-body simulation
distribution as follows:

\begin{enumerate}

\item An observer, situated at an arbitrary position in the simulation
box,
carves out a sphere of radius $r_{\rm max} = 240
\hmpc$.

\item Mass points are selected according the PSCz selection
function:
\begin{equation}
\phi(\Delta) = \phi_{\star} \frac{ 10^{(1 - \alpha)\Delta} }{ \left(1 +
10^{\gamma \Delta} \right)^{\frac{\eta}{\gamma}}},
\label{sf}
\end{equation}
where $\Delta=\log_{10}(r/r_0)$ and the values of the parameters 
are $\phi_{\star} = 0.006794\,h^3$Mpc$^{-3}$, $r_{0} = 1.95113 
\hmpc$, $\alpha = 1.88654$, $\eta =
4.38595$ and $\gamma = 1.54554$. These parameters were derived using 
the methods described by \scite{MST}.

\item Mass points are moved to their redshift space positions,
$\s$, via the equation
\begin{equation}
	\s = \r\left[1 + \frac{(\v-\v_0) \cdot \hat{\r}}{H_{0}r}\right]
\end{equation}
where we have full information on the
velocity $\v$ of each particle.  The observer velocity $\v_0$ is taken to
be zero for the simulations, but is more correctly the Local Group velocity
in the survey (so the second term in brackets does not formally diverge as 
$r\rightarrow 0$). 
\item Cartesian $x$, $y$, $z$ coordinates are converted to galactic
latitude and longitude and a PSCz sky mask (see Section 4.1) is
applied to the distribution.
\item Mass points are drawn at random to have the correct selection function
(\ref{sf}).
\item Mass points in the masked region of sky are discarded.  The mask used 
here is in fact slightly less conservative than that finally used for
the real data.
\end{enumerate}


\begin{figure*}
\centering
\begin{picture}(400,400)
\includegraphics{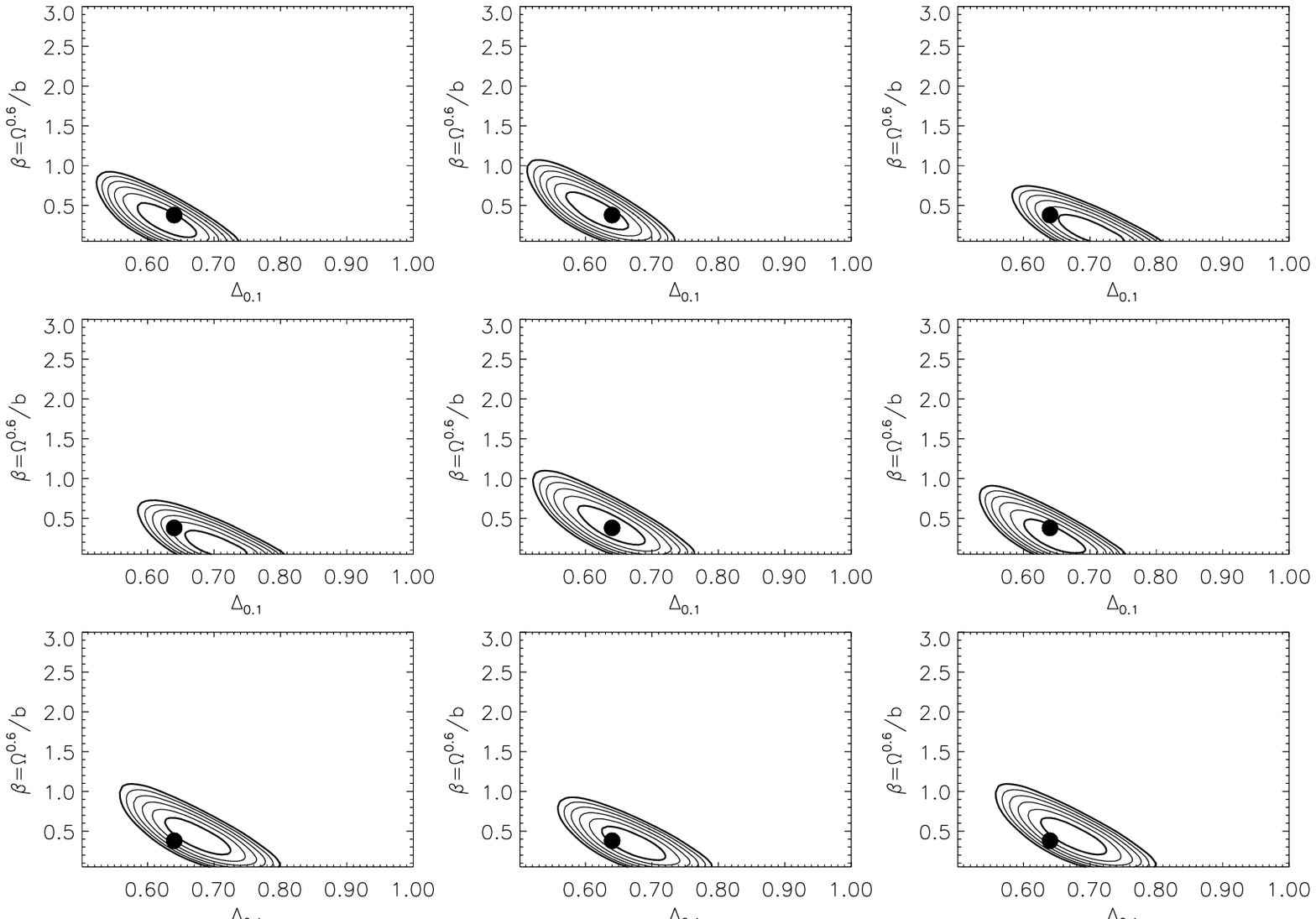}
\end{picture}
\caption[]{\label{LCDM} Contours of likelihood in the
$\beta$ -- $\Delta_{0.1}$ plane for 9 realizations of the PSCz
redshift survey drawn from $\Lambda$CDM N-body simulations. Again, as in all
the following figures, contours
are plotted at intervals of $\delta \ln {\cal{L}} = 0.5$. Points show the
true parameter values.}
\end{figure*}

Using this procedure we obtain mock PSCz surveys with approximately
the same density (and therefore the same number) of mass points as
there are galaxies in the PSCz survey to a proper distance of $r_{\rm max}
\hmpc$. Mass points are taken to represent galaxies throughout the analysis,
so the linear bias parameter is unity. We have made no attempt to
choose `galaxies' from the simulation density distribution as this
introduces extra complications and is unnecessary for the statistical
testing purposes for which we employ the mock catalogues.

\subsection{Results of analysing mock PSCz catalogues}

\begin{figure*}
\centering
\begin{picture}(400,400)
\includegraphics{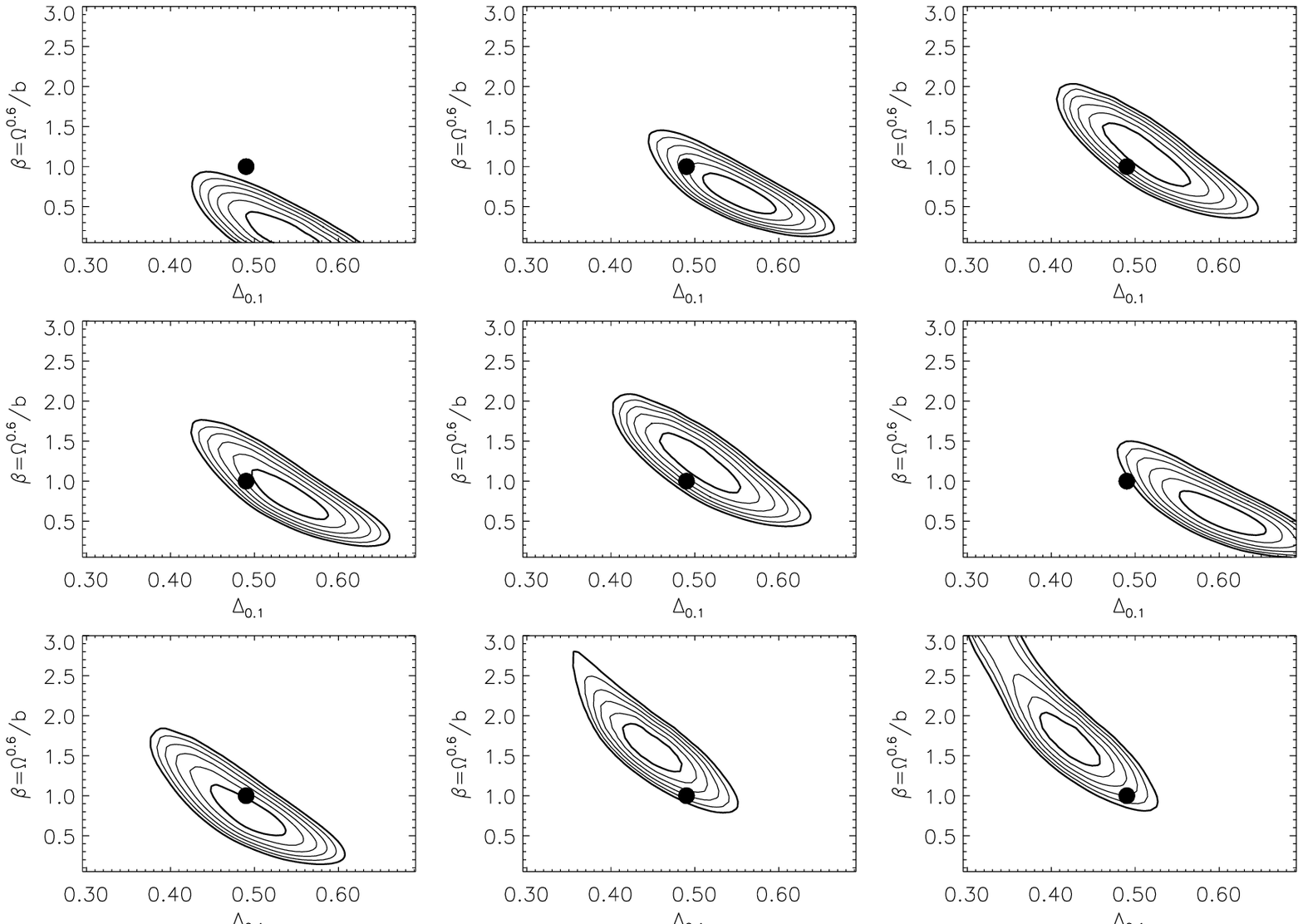}
\end{picture}
\caption[]{\label{MDM} Figure shows contours of likelihood in the
$\beta$ -- $\Delta_{0.1}$ plane for 9 realizations of the PSCz
redshift survey drawn from MDM N-body simulations. }
\end{figure*}

Figures~\ref{SCDM}, ~\ref{LCDM} and ~\ref{MDM} show the result of applying
the analysis described in Section 2.1 to the mock PSCz catalogues. In
all cases, we have used $r_{\rm max} = 240 \hmpc$ and have analysed modes
up to $k = 0.1 \hmpcrev$. The maximum $\ell$ and $n$ values are $21$ and
$8$. For the convolutions we have included modes up to $\ell = 30$, $n =
20$.

For the calculation of the scattering matrix (equation \ref{Smatrix},
Appendix A), 
the three-dimensional velocity dispersion $\sigma_v$ was measured from
the simulations for each model. We have used the values: $\sigma_v =
1160 \kms, \; 587 \kms$ and $859 \kms$ for the SCDM, $\Lambda$CDM and MDM
models respectively. 

In the analysis of the CDM-like mock catalogues we 
have assumed a real-space
power spectrum of the form given in equation (\ref{CDM}). For the MDM model,
a power spectrum of the form of equation (1) of \scite{KHPR93} was
used. These power spectra are needed for the calculation of the
covariance matrix (equations 52 -- 54, Appendix A) as well as for the weighting
function (equation \ref{weight}).

\begin{figure}
\centering
\begin{picture}(200,200)
\includegraphics{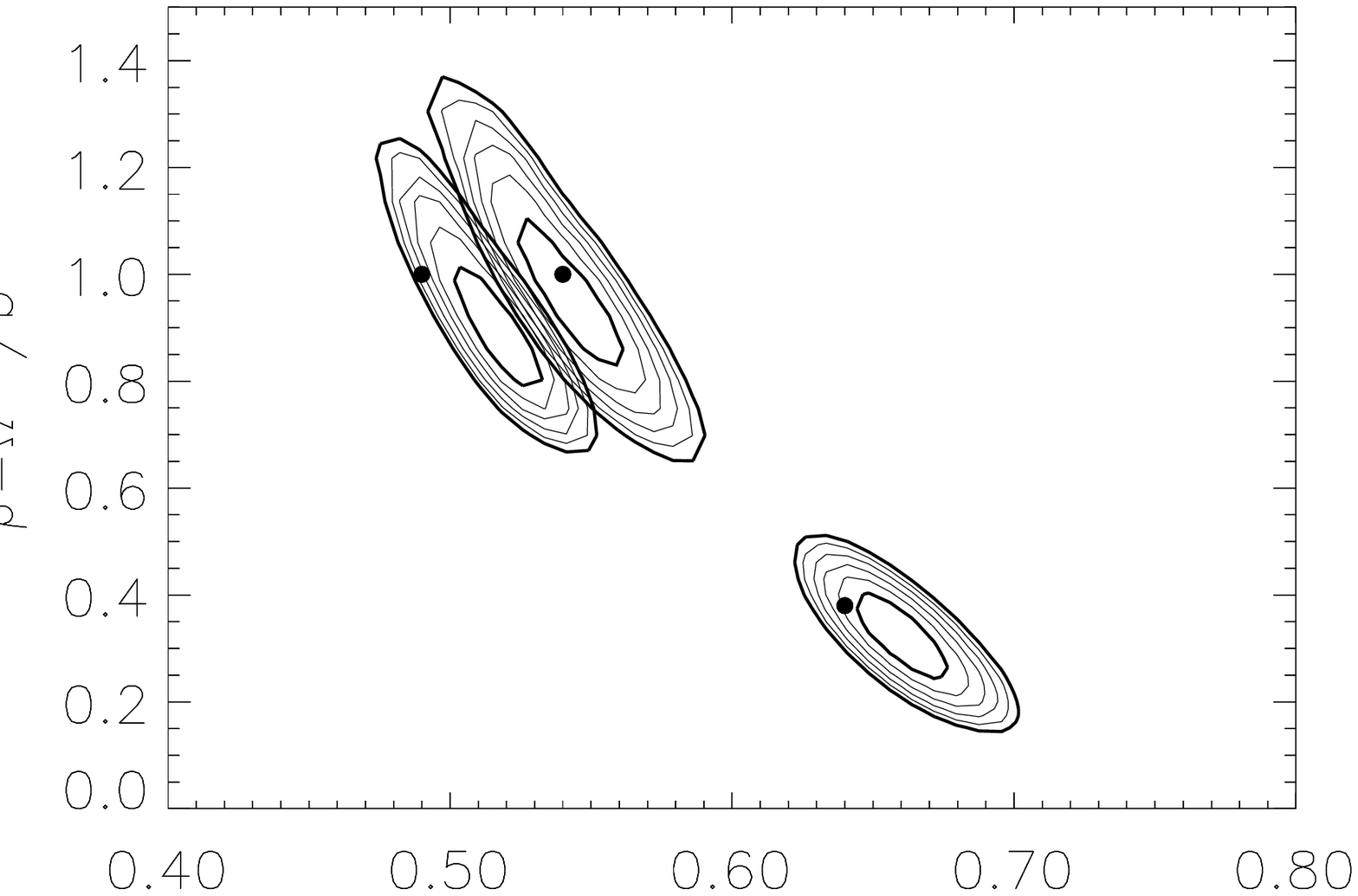}
\end{picture}
\caption[]{\label{model_ens} Parameter likelihoods for ensemble-averaged 
mock PSCz redshift surveys drawn from SCDM simulations (top right),
$\Lambda$CDM simulations (bottom right) and MDM simulations (top left).
True parameter values are marked with a point.
\label{simsens}}
\end{figure}

Figures~\ref{SCDM}, ~\ref{LCDM} and ~\ref{MDM} show contours of likelihood in 
the $\beta - \Delta_{0.1}$ plane for the SCDM, $\Lambda$CDM and MDM mock 
catalogues 
respectively. The contours are plotted at intervals of $0.5$ in log--likelihood
and the $x$-axis is labelled at intervals of $0.1$ in
$\Delta_{0.1}$. The true values of $\beta$ and
$\Delta_{0.1}$ are indicated by points.  Figure \ref{simsens} shows the 
ensemble average of the individual likelihoods and demonstrates the correct 
recovery of the parameters 

The average recovered value of $\beta$ and $\Delta_{0.1}$ over the
nine realizations of the PSCz survey for each of the cosmological
models, together with the true values of these parameters are
presented in Table~\ref{simresults}. The error on the recovered value
of $\beta$ and $\Delta_{0.1}$ is the error on the mean value
(obtained from averaging the estimates, rather than from the ensemble
averaged likelihood plots
of Fig \ref{simsens}) multiplied by $\surd{9}$  i.e. this is the 
error appropriate to the
recovered values of $\beta$ and $\Delta_{0.1}$ from a single
realization of the PSCz survey. This error bar provides us with an
`external' estimate of the error.  The formal `internal' errors on the
values of $\beta$ and $\Delta_{0.1}$ are given by the projection of
$\delta \ln {\cal{L}} = -0.5$ onto the parameter axes, i.e. the
projection of the first plotted contour. Evaluating these points
for each of the plots shown in Figures~\ref{SCDM}, ~\ref{LCDM} and
~\ref{MDM} gives the internal error estimates for $\beta$ and
$\Delta_{0.1}$, together with the one sigma range on this error. These
errors are shown in columns $3$ and $6$ of Table~\ref{simresults} for
$\beta$ and $\Delta_{0.1}$, respectively.  Note that the parameter estimates
for the external error determination were made from likelihood surfaces
which were rather crudely sampled, so the mean of these 
estimates is artificially close to the true solution.


The recovered values of $\beta$ and $\Delta_{0.1}$ are in very good
agreement with the true values. In each model, the correct values of
$\beta$ and $\Delta_{0.1}$ are recovered to within the $1\sigma$
error. The internal error measured from the projection of the first
contour level provides a reasonable estimate of the true error, as can
be seen from the fact that the external and internal error estimates
are in good agreement. The only case where this is not true is for 
$\Delta_{0.1}$ in the
MDM model where the external error appears to be too large. This is possibly due to the fact that these simulations are
not large enough to model the PSCz redshift survey out to a proper
distance of $r_{\rm max} = 240 \hmpc$.

It is remarkable that the correction applied here for the non-linear
effects introduced by the 
small-scale velocity field is adequate even for the SCDM model, which
has a very high three-dimensional velocity dispersion ($\sigma = 1160
\kms$).  Without this correction we would expect the recovered value
of $\beta$ to be significantly lower than the true value as non-linear
(``finger of god'') effects become important even for relatively small 
wavenumbers ($k\sim 0.06 \hmpcrev$ -- see e.g. \pcite{tad96}).

\begin{figure*}
\centering
\begin{picture}(600,350)
\includegraphics{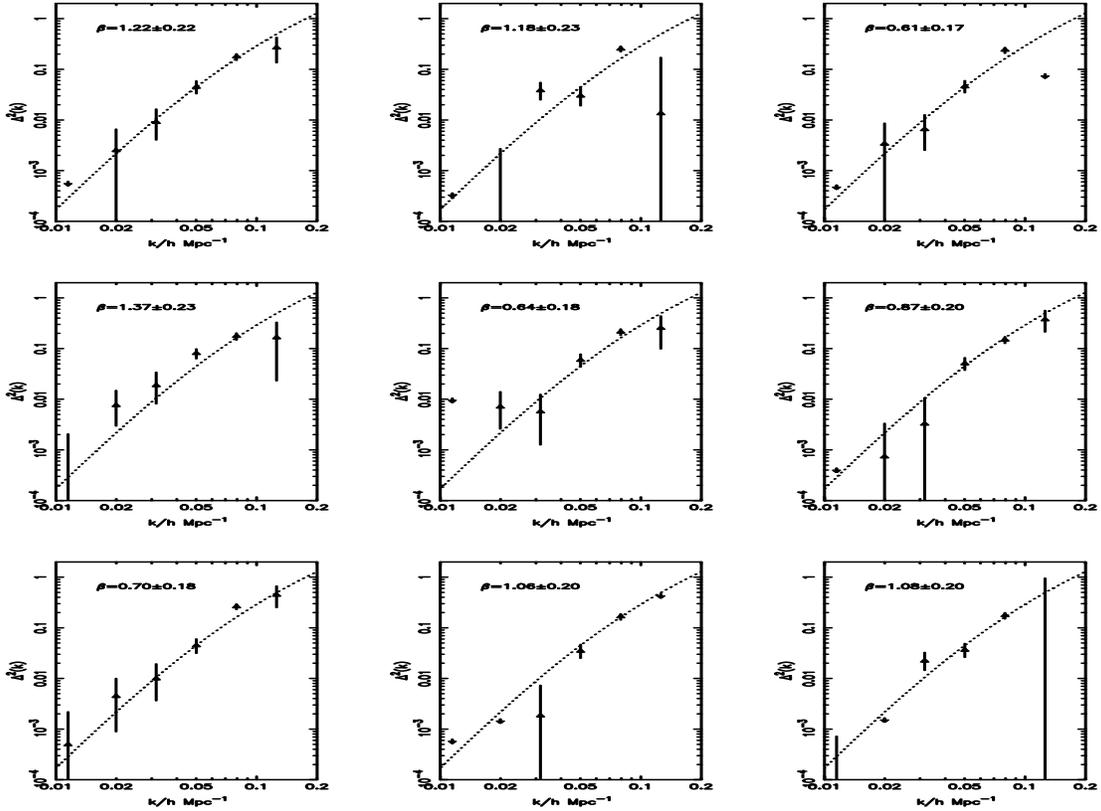}
\end{picture}
\caption[]{\label{LCDM_Pk} The real-space power spectrum 
obtained from an 
ensemble of SCDM mock PSCz simulations (data points), with the true 
power spectrum dotted. For this
analysis, $r_{\rm max}=300 h^{-1}$ Mpc and $k_{\rm max}=0.87 h$ Mpc$^{-1}$.
\label{SCDMPk}}
\end{figure*}

\begin{figure}
\centering
\begin{picture}(400,200)
\includegraphics{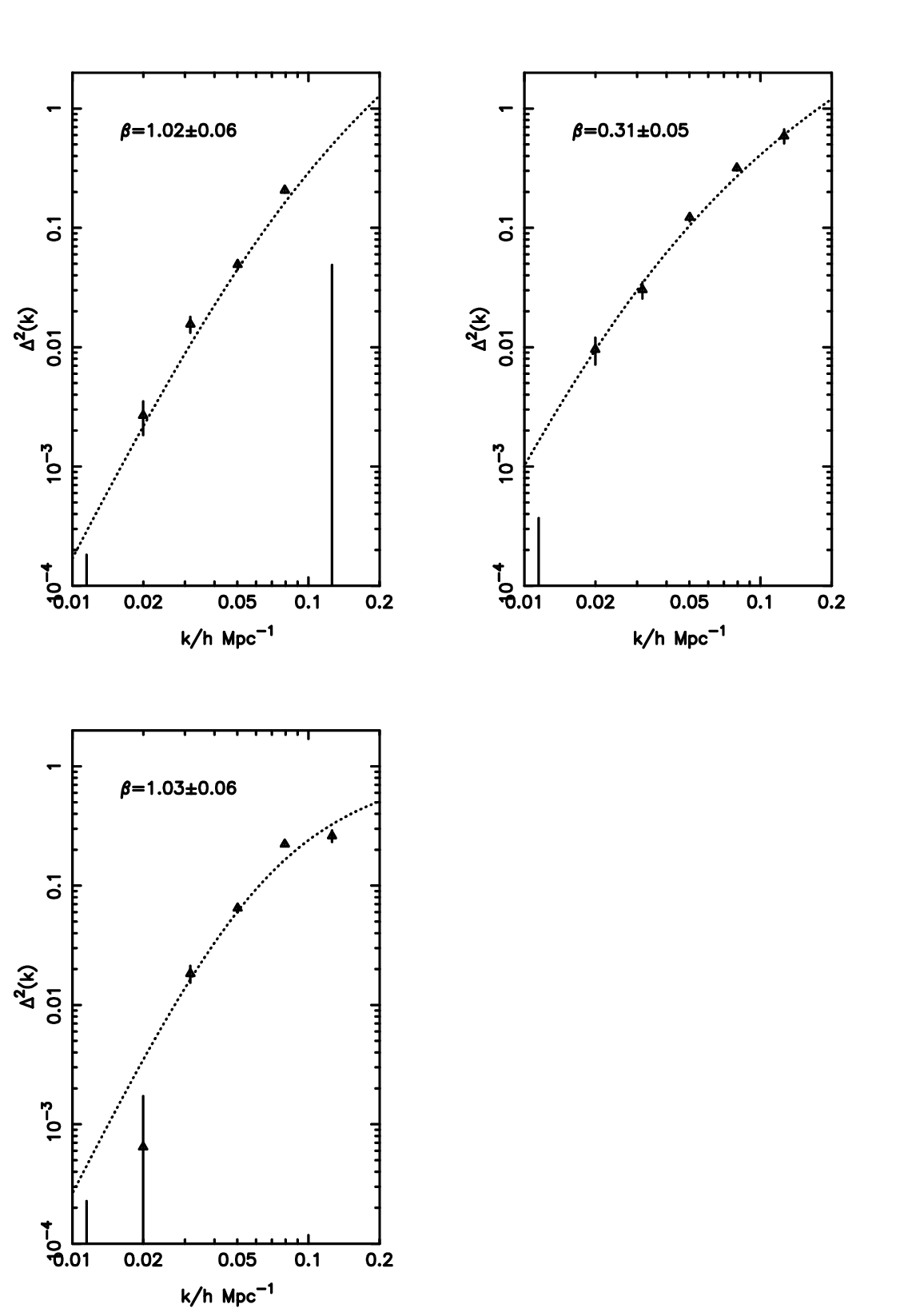}
\end{picture}
\caption{The ensemble average of real-space power spectrum estimates
from the masked SCDM, $\Lambda$CDM and MDM mock surveys, 
along with the true power 
spectrum in each case.  The
recovered value of $\beta$, whose true values are 1.0, 0.38, and 1.0 are 
also shown.
\label{XCDMPk_ens}}
\end{figure}

We have also tested the stepwise maximum likelihood technique using
the mock PSCz surveys described above. 
Figure \ref{SCDMPk} shows the results for individual SCDM 
mock catalogues and Figure \ref{XCDMPk_ens} shows the ensemble averaged 
results for 
all of the SCDM, $\Lambda$CDM and MDM models. As expected in individual 
simulations the uncertainty grows with decreasing wavenumber, but the 
ensemble averaged results show that the stepwise maximum likelihood
method is unbiased. 

Note in particular that the longest measured 
wavelength bin shows no significant bias, since we do not include
the $\ell=0$ mode. This has caused problems in the past for power
spectrum analysis where the mean of the survey is estimated from the survey
itself, suppressing the longest wavelength modes (\pcite{tad96}).

From these tests on simulated PSCz redshift surveys we can be
confident that the analysis described in this paper provides an
unbiased estimate of the parameters $\beta$ and
$\Delta_{0.1}$. Furthermore, the errors obtained by examining the
likelihood contours provide a reasonable estimate of the true error on
$\beta$ and $\Delta_{0.1}$.   

\begin{table*}
{\centerline{\bf Table I}}
\centering
\begin{tabular}{||c|c|c|c|c|c|c||}
\hline 
 Model & $1. \beta_{rec}$  & $2. Error_I$ & $3. \beta_{true}$ & 
 $4. \Delta_{rec}$ &
 $5. Error_I$ &  $6. \Delta_{true}$  \\ \hline \hline
SCDM & 1.01 $\pm$ 0.24 &  0.17 $\pm$ 0.05& 1.00 &  0.53 $\pm$ 0.03 &
 0.04 $\pm$ 0.01 & 0.54 \\ 
$\Lambda$CDM & 0.36 $\pm$ 0.12 & 0.15 $\pm$ 0.04 & 0.38 &  0.66 $\pm$ 0.03 &
 0.05 $\pm$ 0.01& 0.64\\ 
MDM  & 1.00 $\pm$ 0.60 & 0.20 $\pm$ 0.05 & 1.00 &  0.51 $\pm$ 0.06 &
 0.03 $\pm$ 0.01 & 0.49  \\ \hline
\end{tabular}    
\caption{\label{simresults} The results of applying the analysis described
in Section 2.1 to mock PSCz redshift surveys drawn from three cosmological
models. Columns $1$ and $4$ give the recovered values of $\beta$ and
$\Delta_{0.1}$, averaged over nine realizations, with errors derived
from the scatter in the recovered values (external errors). These are to be
compared with the true values of the parameters, given in columns $3$
and $6$. Columns 2 \& 5 show the mean and scatter of the 1-$\sigma$ projected
error bars (internal errors).} 
\end{table*}


\section{The IRAS PSCz redshift survey}
We are now in a position to apply the analysis described above to the
newly completed IRAS PSCz redshift survey. In Section 4.1 we briefly describe
the survey itself, and in Sections 4.2 \&
4.3  we test for various systematic effects in the catalogue that could 
bias the recovered value of $\beta$. The results of our analysis are 
presented in Section 5.

\subsection{The Dataset}
\begin{figure}
\centering
\begin{picture}(200,200)
\includegraphics{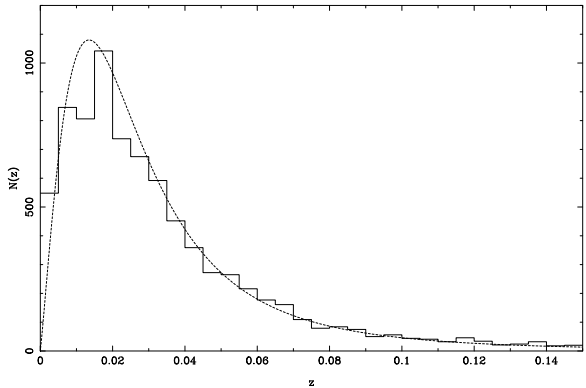}
\end{picture}
\caption[]{\label{nzpscz} The redshift distribution for the galaxies in the 
$>$0.75Jy PSCz sample, in the unmasked regions of sky.  For the analysis, the 
data are also cut at various values of $z_{max}$ around 0.08.  The line 
shows the selection function used.}
\end{figure}

The IRAS PSCz survey is the largest and deepest all sky redshift
survey of galaxies to date. It was conducted as a follow up to the $1$
in $6$ sparsely sampled QDOT redshift survey (\pcite{QDOT}). The aim
was to determine redshifts for all galaxies in the IRAS Point Source
Catalogue (PSC -- \pcite{psc}) to a 60$\mu m$ flux limit of 0.6Jy in
the area of sky where reliable optical identifications and redshift
acquisition were feasible. An outline of survey construction is given
in \scite{willherst}. About $4600$ new redshifts were measured in the
course of the project, the rest being taken from the literature or
obtained via private communications. Details of the data reduction process will be published elsewhere -- (Keeble \et in preparation).  Figure~\ref{nzpscz} 
shows the $N(z)$ distribution
for galaxies in the PSCz redshift survey. The median redshift of galaxies 
in the
survey is $\sim 0.03$, but there is a tail of galaxies out to
redshifts as large as $\sim 0.2$. The solid curve shows the predicted
$N(z)$ using the selection function of equation (\ref{sf}).
The total error on the measured velocities is $\sim 100$
\---\ $150 \kms$, which is negligible for this analysis.

The PSCz was undertaken for various analyses, some of which demand the 
largest sky coverage possible. The default PSCz mask defining the survey consists of the IRAS coverage gaps, the LMC and SMC, and areas estimated on the basis of their $100\mu m$ sky brightness (Rowan-Robinson \etal 1986) to have $A_B > 2^m$, altogether 16\% of the sky (\pcite{SPSCZ98}). At low latitudes there is inevitable 
degradation in the survey due to incomplete identifications and the greater difficulty of getting redshifts. For an analysis such as this, where we are looking for weak clustering over large scales, uniformity is paramount. We therefore created more conservative masks, corresponding to optical extinctions of $A_B=0.5^m,0.75^m,1^m$, for use in this analysis. The results presented are all for $A_B=0.75^m$ (excluding 35\% of the sky), but are robust to which of these three latter masks is used. 

There is also incompleteness at high redshift, in that galaxies with than $A_{bJ} > 19^m$ were not systematically pursued spectroscopically. Based on the joint optical/$60 \mu m$ luminosity functions of \scite{S90} and the ULIRG survey of \scite{Clem96}, we believe that a negligible number of these faint galaxies are closer than $z=0.1$, for areas with $A_B<1^m$.

Within the volume defined by these cuts in area and distance, there are potentially variations in the IRAS data quality due to: the number of scans made by the satellite (HCONs), the South Atlantic Anomaly, Malmquist effects, hysteresis, confusion and noise lagging (\pcite{psc}). The most serious of these is the first, causing known incompleteness at faint fluxes. We restored most of this by reclaiming galaxies from the PSC reject file (\pcite{SPSCZ98}). We investigated the residual quality variations both theoretically and empirically (through source counts), and were unable to find evidence for any non-uniformities above a level of 2\% in flux or 3\% in density. Several effects are expected to come in below this level, and so these numbers represent our overall best-guess estimates of the PSCz uniformity. Most of the concerns about the uniformity of the PSC affect sources near our flux limit of 0.6Jy, and so our final quoted results refer to a sample cut conservatively at 0.75Jy. 

\begin{figure*}
\begin{center}
\setlength{\unitlength}{1mm}
\begin{picture}(90,70)
\includegraphics{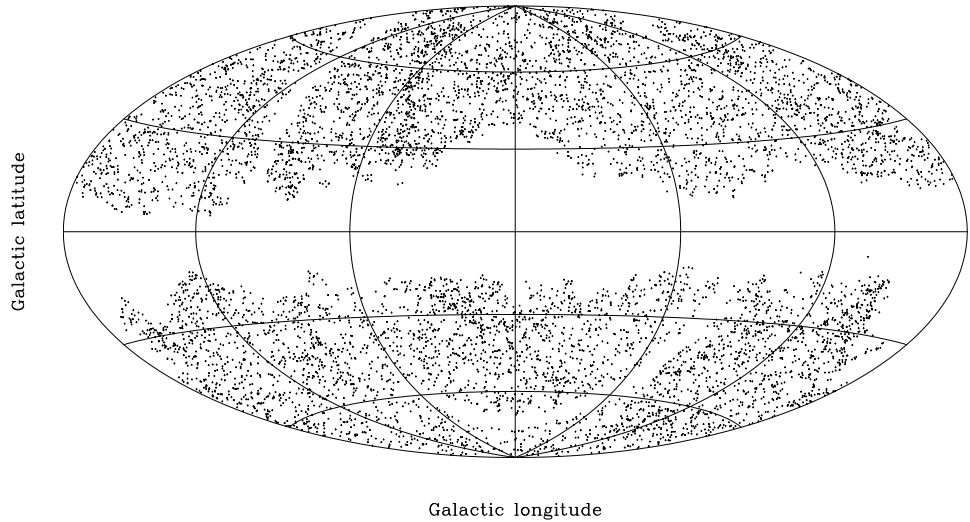}
\end{picture}
\end{center}
\caption[]{\label{psczsky} The sky-distribution of galaxies in the
PSCz survey with 60$\mu m$ flux above 0.75Jy. Here we have applied the
conservative sky-mask used in our analysis.  No cut in redshift has
been imposed.}
\end{figure*}

The distribution of the PSCz
galaxies used in this analysis on the sky is shown in Figure~\ref{psczsky}.

\subsection{Internal consistency checks}

The ability to make high-precision estimates of the power spectrum
with this method allows us to look for systematic effects in the catalogue.
We do this by making a number of changes in the sample and the modes
analysed and looking for inconsistencies in
the recovered power spectrum and $\beta$. In this section, we give details 
of the various tests for systematics we have applied to the PSCz,
and in  Section \ref{inconres} we describe the results.

We have tested the catalogue (and the method) for sensitivity to the
following:
\begin{itemize}
\item Changes in the upper wavenumber limit, $k_{\rm max}$
\item Changes to the outer boundary, $r_{\rm max}$
\item Evidence for excess angular variance (explained below)
\item Changes to the flux limit
\item Changes from point source fluxes to addscan fluxes
\item Changes between a high flux sample and low flux sample.
\end{itemize}

By 'addscan' fluxes, we mean fluxes determined from the coadded raw IRAS data (kindly provided by IPAC), using software provide by Amos Yahil.
The last test mentioned above is an important one, and needs further
explanation.  We first divide the PSCz catalogue into thin radial
shells (width 100 km s$^{-1}$). The median flux of galaxies in a shell
is calculated and galaxies with flux above this value are put into a
`bright' catalogue (the high-split catalogue) and the galaxies with
flux below the median go into the low-split catalogue. This
effectively divides the PSCz by luminosity while ensuring that the two
sub-samples have the same radial distribution. Consequently, any
inconsistencies between the results from the high and low-split
catalogues cannot be due simply to changes in selection function or,
equivalently, the radial distribution of points.  We should point out
that demonstrating consistency is not always straightforward: even
disjoint samples of galaxies are not independent, as they are (by
assumption) drawn from the same underlying density field.  They do,
however, have independent shot noise.

\subsection{Results of consistency tests}
\label{inconres}

For speed, we generally chose a relatively small wavenumber cutoff, 
$k_{\rm max}=0.08 h$ Mpc$^{-1}$, when performing these consistency tests.
We find no evidence for any systematic changes in the results for changes 
in $r_{\rm max}$, or $k_{\rm max}$, or from changing point source fluxes 
to addscan fluxes. 
However when we changed the flux limit of the PSCz sample from 0.6Jy
to 0.75Jy the results for $\Delta_{0.1}$ from these
catalogues (hereafter PSCz0.6 and PSCz0.75) differed
significantly. The high/low-split catalogue results show this
inconsistency more clearly as explained in the next section.

\begin{figure}
\centering
\begin{picture}(200,200)
\includegraphics{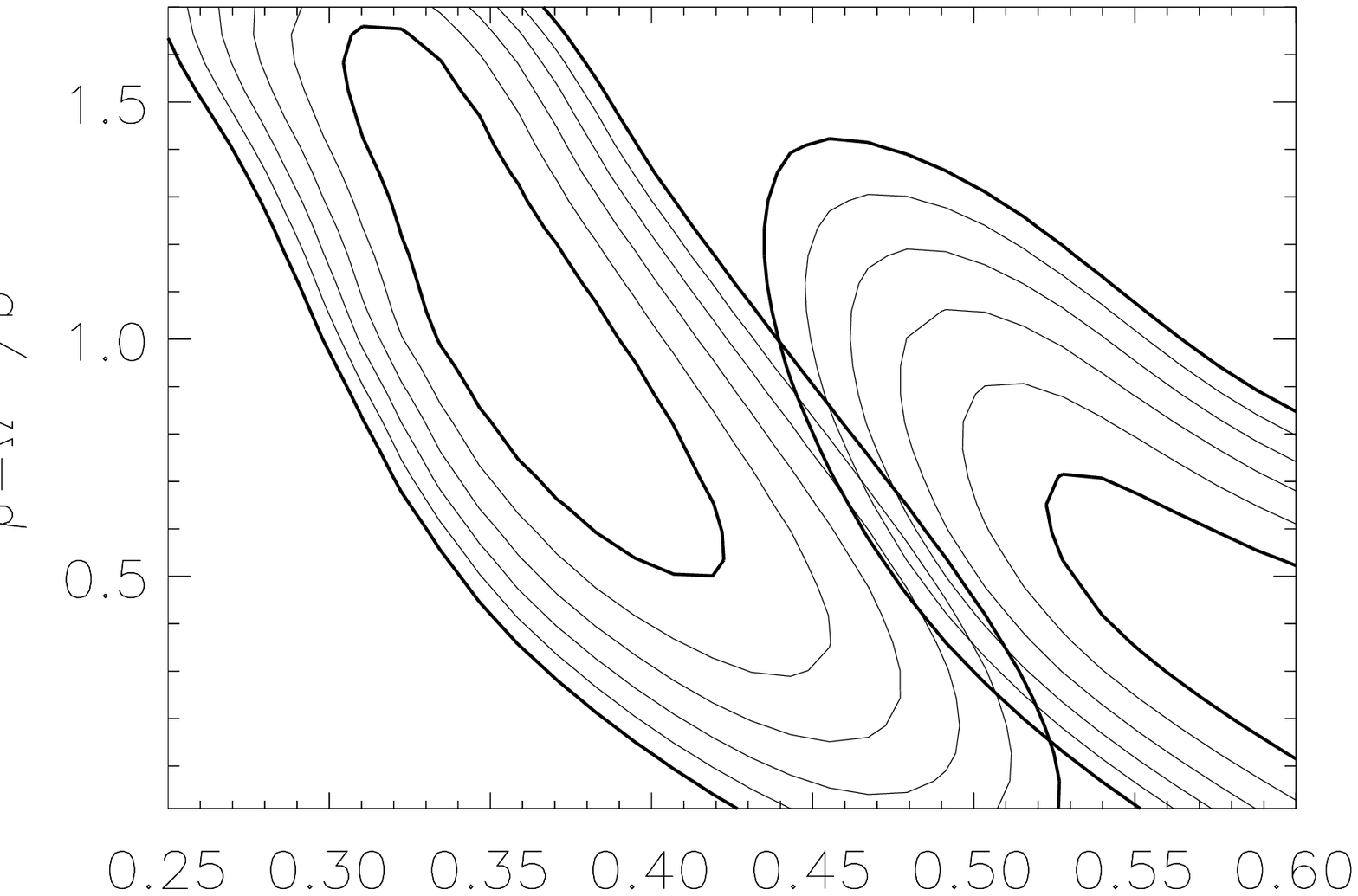}
\end{picture}
\caption[]{Likelihood contours for the two subsamples of PSCz0.6
(galaxies with $60\mu$ flux about 0.6Jy) split by luminosity. Here we
have used a restricted set of modes, $k_{max}=0.08$, and a boundary at
200 $h^{-1}$ Mpc. The high and low luminosity subsamples are created
as described in the text. Contours for the low luminosity sub-sample
are to the bottom right of the plot. The amplitude, $\Delta_{0.1}$,
is seen to be significantly higher for the low luminosity sub-sample.}
\label{IC6}
\end{figure}

\begin{figure}
\centering
\begin{picture}(200,200)
\includegraphics{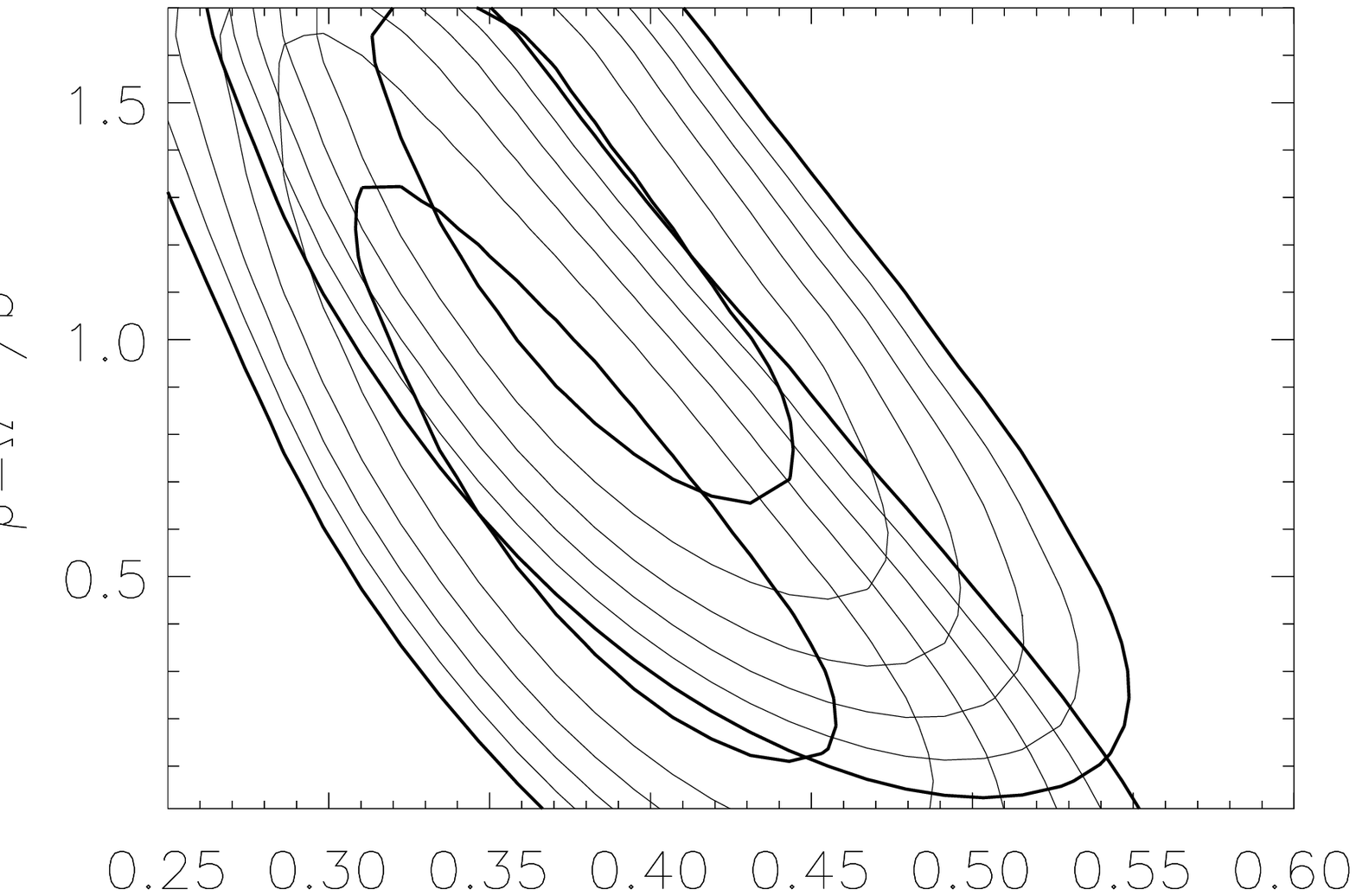}
\end{picture}
\caption[]{Likelihood contours for the two subsamples of PSCz0.75
(galaxies with $60\mu$ flux about 0.75Jy) split by luminosity. Again, we
have used a restricted set of modes, $k_{max}=0.08$, and a boundary at
200 $h^{-1}$ Mpc. It can be seen that by excluding the galaxies in PSCz
with 60$\mu m$ flux between 0.6Jy and 0.75Jy before splitting the
sample by luminosity we obtain parameter estimates for the high/low
luminosity subsamples which are consistent. }
\label{IC745}
\end{figure}

We can make high and low-split catalogues from both PSCz0.6 and
PSCz0.75. Figures \ref{IC6} and \ref{IC745} show parameter estimates for
the high-split and low-split catalogues for these two flux limits.
Figure \ref{IC6} shows clear problems for the PSCz0.6 sample, as
evidenced by the inconsistent parameter estimation from the high and
low split subsamples.  PSCz0.75 on the other hand, gives consistent
results.  We have performed a counts-in-cells analysis (c.f
\pcite{GPE95}), and similarly find an excess variance in the low-flux
data. 

The PSC has known imperfections at our flux limit of 0.6Jy. These
include incompleteness in 2HCON areas, a large and crude `faint flux
overestimation' correction factor, and low and variable signal-to-noise 
(all \pcite{Beich}).  These problems have been investigated and,
where possible corrected for in the galaxy catalogue used for this
survey (\pcite{SPSCZ98}).  Nevertheless, the completeness and
uniformity, which are paramount to this analysis, are certain to
deteriorate towards the flux limit. 

Because we find that our results are sensitive to the flux limit we
use we have conservatively chosen to limit the catalogue to 0.75Jy in
our quoted results.

However, despite extensive tests, we have not been able to find any `smoking
gun' in the sense of identifying where the excess clustering is coming
from at lower fluxes.  In fact, PSCz0.6 gives a result for $\beta$
which is consistent with the PSCz0.75 sample; it is the amplitude
which is the problem, being enhanced over PSCz0.75 by 15\%.  Given the
smallness of the errors in the amplitude such a discrepancy is
significant.

We have also searched for, and failed to find, a signature of
direction-dependent flux errors, as proposed by \scite{hameff}.  This
arose from his analysis of the redshift-space correlation function of
various IRAS surveys.  He found an elongation along the line-of-sight
on larger scales than plausibly accounted for by fingers-of-god.  This
was particularly marked for the QDOT survey in the region beyond 80
$\hmpc$, and one explanation proposed was a direction-dependent flux
error of $\sim 0.1 Jy$, varying on angular scales of $7^{\circ}$.

We can check partly for this effect by inspecting the $n=1$ modes for
signs of excess angular variance.  These modes, which oscillate once
radially, should be enhanced, as flux errors would affect the distant
part of the catalogue more than the nearby region.  We see no such
signature: the power in these modes is consistent with the model
solution and shot noise.  It should be stressed however that our
large-scale analysis does not probe angular scales as small as
$7^{\circ}$ ($\ell=50$); it is only sensitive to variations on rather larger
angular scales. 

We also rule out a straightforward radially-dependent effect
as the cause of the discrepancy between the high and low luminosity
sub-samples of PSCz0.6, since these sub-samples have the same mean radial
distributions by construction. \scite{Suth98} have
analysed the redshift-space power spectrum and find the same behaviour
as seen here, i.e. a higher amplitude when all galaxies down to the
0.6 Jy flux limit are included.  The effect is smaller in redshift
space, but this would be expected if the lower flux
galaxies had an excess variance, for whatever reason.

If it is a
real luminosity effect (i.e. faint galaxies cluster more strongly than
bright galaxies, introducing a differential bias), it must be very large, 
since in the analysis, the
samples with different flux limits differ little in the weight given
to galaxies of different luminosity. Moreover, if it is a true
luminosity effect we would also expect to see it in PSCz0.75.

In our desire to leave no stone unturned, we also investigated the
selection function itself, as the method of determining it
\cite{MST} provides the selection function in redshift space, not real 
space.  The effects of this can be accounted for through a Taylor
expansion and the addition of an extra term in the equations, but
inclusion of this extra term makes an infinitesimal difference to the
parameter estimation, so we neglect it.

\section{The redshift distortion of the PSCz and the amplitude of 
the real space power spectrum}

We have analysed the PSCz0.75 survey using the method described in Section
2.1. Here we present results for a boundary at $r_{\rm max} = 200 \hmpc$.   
With the 
sky mask as defined in Section 4.1 we have $7042$ galaxies in the data-set 
to the flux limit of $0.75$Jy.  
We use a wavenumber limit of $k_{\rm max} = 0.13 h$ Mpc$^{-1}$.  These modes 
have $\ell$ up to 25, and $n$ up to 9, but not all of the modes with these 
limits are used, as many fall outside the wavenumber range.
We include modes up to $\ell = 30$ and $n = 20$ for the convolutions.

For the calculation of the covariance matrices we have assumed a
CDM-like real space power spectrum with $\Gamma = 0.2$ (equation \ref{CDM}). 
As explained in Section 2.1, although the shape of the real space
power spectrum in this analysis is fixed, the amplitude at a
wavenumber $k = 0.1 \hmpcrev$ is the second free parameter in the
likelihood fit. A wavenumber-dependent weighting function (equation
\ref{weight}) has been employed. Again, we have used a CDM-like power spectrum
with $\Gamma = 0.2$ for $P_w(k)$. The normalization of $P_w(k)$ is
such that the amplitude $\Delta^{2}_{0.1}$ is 0.18. In this minimum
variance weighting scheme, the exact form
and amplitude of the function $P_{w}(k)$ will not be critical. 

We have employed a value of $\sigma_v = 420 \kms$ for the
three-dimensional velocity dispersion of galaxies in order to
calculate the scattering matrices and correct for non-linearities
(equation \ref{Smatrix}).
This is the value of $\sigma_v$ obtained by adopting a pairwise velocity
dispersion of  $340 \kms$ (\pcite{huchracfa},
\pcite{FDSYH94}) and assuming that the small scale peculiar
velocity field is incoherent. 

\begin{figure}
\centering
\begin{picture}(200,200)
\includegraphics{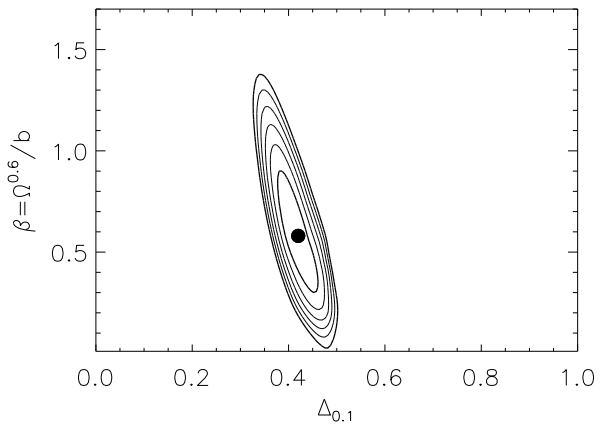}
\end{picture}
\caption[]{\label{results1} Likelihood contours for the two parameters
$\beta \equiv \Omega_0^{0.6}/b$ (where $b$ is the IRAS bias parameter)
and the amplitude of the power spectrum (parametrised by $\Delta^2(k)
\equiv k^3 P(k)/(2\pi^2)$ at a wavenumber $k=0.1\,h$ Mpc$^{-1}$) for
the PSCz survey flux-limited to 0.75Jy.  The outer boundary is 200
$h^{-1}$ Mpc, and all wavenumbers up to $k=0.13 h$ Mpc$^{-1}$ have
been analysed.  A conservative mask with extinction $A_B < 0.75$ has
been applied, leaving a total of 7042 galaxies in the sample.}
\end{figure}

Figure~\ref{results1} shows the contours of likelihood in the $\beta -
\Delta_{0.1}$ plane, plotted at intervals $\delta \ln {\cal L} = 0.5$
for the PSCz survey. The recovered values of $\beta$ and
$\Delta_{0.1}$ are:

\begin{eqnarray*}
\beta = 0.58\pm 0.26\nonumber \\
\mbox{\hspace{1cm}}\Delta_{0.1} = 0.42\pm 0.03.\\ \nonumber
\end{eqnarray*}	
We have analysed PSCz0.6 to a wavenumber limit of 0.12 $h$ Mpc$^{-1}$;
this gives $\beta = 0.57\pm 0.25$ but $\Delta_{0.1} = 0.48\pm 0.04$.
However, as mentioned earlier, there are inconsistencies in detail which 
lead us not to trust the results from the full 0.6 Jy sample.

\subsection{The shape and amplitude of the  real--space power 
spectrum of IRAS galaxies}

\begin{figure}
\centering
\begin{picture}(200,240)
\includegraphics{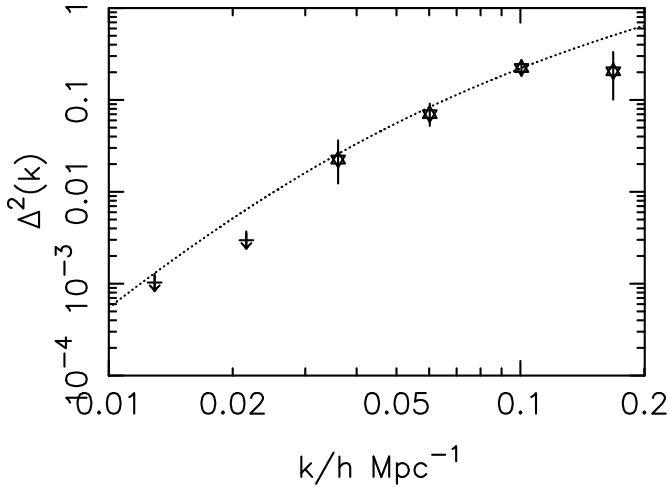}
\end{picture}
\caption{
The real-space power spectrum of the PSCz redshift survey in dimensionless 
units. The curve is a CDM model with $\Gamma = 0.2$.  The estimated 
redshift distortion is $\beta = 0.47 \pm 0.16$ (conditional error).}
\label{Pk}
\end{figure}

The stepwise power spectrum was fitted as described in Section 2.2 and 
has been applied to the 1.2Jy survey (BHT). Unlike the BHT analysis, we 
have chosen 
that the right hand bin  should begin at $k_{\rm max}$, to show how 
information beyond the observed 
modes can be analysed. There is a constraint on this bin due to the 
effect of mixing, but this is quite weak as can be seen by the large 
error bar on the final point in Figure \ref{Pk}.
Again, reassuringly, the redshift distortion result is similar to that 
of the two-parameter fit with
\be
\beta = 0.47 \pm 0.16,
\ee
although note that this is a conditional error, as are the error bars 
on the power spectrum.  The analysis has $r_{\rm max}=200 \hmpc$ and
$k_{\rm max}=0.13 h$ Mpc$^{-1}$.  Reducing the wavenumber limit to 0.12 
gives a consistent result of $\beta = 0.60 \pm 0.18$.

The power spectrum again fits low $\Gamma$ CDM models very well, 
although there is some evidence that the break is slightly sharper 
as suggested by Peacock (1997). It is perhaps possible that the final,
high-$k$ point is suppressed by nonlinearities, although the random smoothing 
model was used.
A particular point to note is the very tight lower limit on the first 
two bins, close to the value expected in current models. This demonstrates 
that the sample we finally use after the cuts discussed in Section 4.3 is 
uniform (in the sense of rms variations) to better than $3\%$ on scales of 
$500\hmpc$.

\section{Discussion and Conclusions}

We have presented the results of a likelihood analysis of a spherical
harmonic expansion of the PSCz galaxy survey.  The aim has been to
establish both the real-space power spectrum and the degree of
redshift distortion from the distribution of galaxies in redshift
space.  The main assumptions in the analysis are that structure grows
via gravitational instability, and that the galaxies are linearly
biased tracers of the underlying mass distribution, at least on the
scales which we analyse (wavelengths $>48 h^{-1}$ Mpc).  There are
many attractive features about this analysis, and it is worth
reiterating them here.  Firstly, the expansion in spherical polar
coordinates is ideal for many redshift surveys, but particularly for
the IRAS galaxy surveys, which cover a large fraction of the sky.  The
effects of redshift distortion are radial, so are naturally accounted
for in this system; in addition the masked regions of sky, due to the
galactic plane, unobserved regions of sky, and patches badly affected
by cirrus, are specified in terms of the angular coordinates alone.
These facts allow a relatively simple translation from real- to
redshift-space expansion coefficients in linear theory.  This is a
major advantage for relatively shallow IRAS surveys, since the linear
modes typically subtend large angular scales.  A general feature of
redshift distortion studies to obtain $\beta$ is that all the
information is obtained from the galaxy positions, rather than relying
on peculiar velocity measurements. The particular expansion we have
used here is effectively a power spectrum analysis in spherical
coordinates.  Because of the survey geometry and redshift distortions
we do not measure the power spectrum directly, but the window
functions converting real-space power to redshift-space expansion
coefficients are narrow, and we are able to measure with some accuracy
the power spectrum on large scales between 40 and 270 $h^{-1}$ Mpc. In
addition constraints on the power can be put on scales up to $\sim 500
\hmpc$.

The analysis we have done follows closely the work of \scite{HT} and 
\scite{Ballinger}, although we have made some technical improvements in the
treatment of the mask and the local group motion, but the effects of the
approximations in the previous papers are in fact negligibly small.

The results we present are conservative in a number of ways.  We have
firstly applied a rather conservative mask, and consider only regions of
sky where the extinction $A_B<0.75$.  Secondly,
we found that we were able to obtain robust results only for a subset of 
the catalogue cut at a higher flux than the formal 0.6 Jy limit.  The
most striking example of the inconsistency near the flux limit is illustrated
in Figure \ref{IC6}, where we have taken care to analyse two subsamples 
with the
same radial and angular distribution to minimise other sources of error.
This graph shows why we do not trust the catalogue at the lowest flux limits,
although we have not been able to pin down the source of the discrepancies.
What we do find is that the results are quite stable if we include galaxies
down to a flux limit of 0.75 Jy.  The price we pay for these conservative
choices is that the galaxy sample is cut back to about 7000 galaxies,
and the error bars are larger than we would have liked.  However, we believe 
that the resulting estimates and errors can be trusted.  In the first 
analysis, we assume that the power spectrum has the shape of a $\Gamma=0.2$
CDM spectrum, at least on scales larger than 40 $\hmpc$, and estimate
simultaneously $\beta$ and the amplitude (characterised by the 
dimensionless power at 0.1 $\hmpc$).  This analysis yields
\ba
\beta & = & 0.58 \pm 0.26\nn
\Delta_{0.1} & = & 0.42 \pm 0.03.\nn
\ea
If we allow the power spectrum to be free (or at least characterized by a large
number of parameters), the results change only slightly:
\ba
\beta & = & 0.47 \pm 0.16\nn
\Delta_{0.1} & = & 0.47 \pm 0.03\nn
\ea
but note that in this case the quoted errors are conditional, and are not
reflective of the true marginal errors.

The value of $\beta$ which we derive is lower than the values of 1.1 and 1.0 
obtained by similar analysis of the 
IRAS 1.2 Jy catalogue (\pcite{HT}, \pcite{Ballinger}), and by the analysis,
also using spherical harmonics, of \scite{FSL94}.  However, the errors in our
previous analyses are large enough to allow $\beta$ to be as low as 0.5. 
Our result is consistent with determinations based on the plane-parallel 
approximation (e.g. \pcite{CFW94},\pcite{CFW95}) and correlation function
studies \cite{HAM92}.  A full discussion of results in this area can be
found in \scite{ham97}.   Methods based on comparison between velocity fields
and IRAS density fields also show a (formally inconsistent) 
spread of results, from $0.49 \pm 0.07$
\cite{WSDK97} to $0.89 \pm 0.12$ \cite{SEDSY98}.  One possible way to 
reconcile 
discrepant results in general is if bias is scale-dependent (e.g. 
\pcite{Coles93}, \pcite{Mann98}), and the measurements are effectively probing
different scales, but it is not clear whether this is appropriate here.

Since we compute the power spectrum and the redshift distortion simultaneously,
and assuming the linear bias model,
we are able to express the answer in terms of the strength of clustering of the
mass, as has been done from the abundance of clusters (\pcite{WEF93}).
We express the amplitude of matter clustering in terms of the dimensionless
matter power spectrum at a wavenumber of 0.1 $h$ Mpc$^{-1}$:
\be
\Delta_{0.1, \rho} = 0.24 \Omega_0^{-0.6}
\ee
with an error of about 50\%. 
If we convert this to the r.m.s. fluctuations in an 8 $\hmpc$ sphere, assuming 
a $\Gamma=0.2$ shape parameter (using the approximation equation \ref{s8}), 
we get
$\sigma_8 = 0.38 \Omega_0^{-0.6}$, again with an error of about 50\%.  This
rises to 0.57 if $\Gamma=0.5$.

What should we conclude from this?  The simplest, although not the
only, interpretation is that IRAS galaxies are unbiased, and $\Omega_0
\simeq 0.4$.  The strongest indications that the Universe might be
Einstein-de Sitter have in the past come from the high values of
$\beta$ from peculiar velocity and redshift distortion studies.  The
picture is still not clear, but the evidence is shifting away from a
high $\beta$ towards low.  In any case, it is impossible to make firm
statements about $\Omega_0$ based on these studies because of our
ignorance of the bias parameter, and we shall almost certainly have to
wait until large redshift surveys of $10^5-10^6$ galaxies are
completed, when higher-order statistics will lift the degeneracy
between $\Omega_0$ and $b$ (e.g. \pcite{MVH97}, \pcite{VHMM98}).
Indications of a low value of $\Omega_0$ come from the supernova
Hubble diagram (\pcite{Perl98}, \pcite{Riess98}) combined with rather
uncertain flatness constraints from the microwave background, although
the prospects are good \cite{TEH98}.

\section*{Acknowledgments}

We are grateful to the many people who provided extensive and invaluable
assistance in terms of telescope observations, provision of redshifts,
optical and IRAS information and reduction software. The PSCz used
observations from the INT, AAT and CTIO 1.5m telescopes, had extensive
assistance from the ROE and IPAC, and made great use of the NED, SIMBAD
and ZCAT databases. Computations were carried out on STARLINK facilities.

\bibliographystyle{mnras}
\bibliography{general}

\onecolumn

\section*{Appendix A: Outline of the spherical harmonic formalism}

In this Appendix we outline the spherical harmonic formalism 
used to estimate the power spectrum and redshift space distortion
in the PSCz. The general derivation is similar to that given in HT,
but is somewhat more comprehensive.

\subsection*{A.1 Definitions}
Following HT we expand the density field of the galaxy
distribution in spherical harmonics, $Y_{\ell m}$, and
spherical Bessel functions, $j_{\ell}$,
\be
	\hat{\rho}_{\ell mn} = c_{\ell n} \int \!d^3s \,\rho(\s) w(s)
		j_{\ell}\left(k_{\ell n}s\right)
		Y^*_{\ell m}\left(\theta,\phi\right),
\label{26}
\ee
where $w(s)$ is an adjustable weighting function and $\s$ is the
redshift-space position variable. $c_{\ell n}$ are normalisation
constants, and $k_{\ell n}$ are discrete wavenumbers (see A2).  The
inverse transformation is
\be
	\rho(\s) = \sum_{\ell m n} c_{\ell n} \rho_{\ell mn} j_{\ell}(k_{\ell n} s)
	Y_{\ell m}(\theta,\phi).
\ee
	We use the Binney and Tremaine (1987) definition of spherical harmonics
which uses the Condon--Shortley phase condition;
\be
	Y_{\ell m}(\theta,\phi)  =  \sqrt{ {2\ell+1\over 4 \pi}
{(\ell-|m|)!\over(\ell+|m|)!} } P_\ell^{|m|}(\cos\theta)
 e^{im\phi} \times \left\lbrace {(-1)^m \qquad m\ge
0 \atop 1 \qquad m<0 }\right.
\ee
where $P_\ell^{|m|}(x)$ is the Legendre polynomial defined (for positive
$m$ only) by 
\be
	P_\ell^{m}(x) = (1 - x^2)^{m/2} \frac{d^m}{d x^m} P_\ell (x).
\ee
Note that some definitions add an extra $(-1)^m$ term 
(e.g. Press et al. 1980). With the Binney and Tremaine definition the 
Hermitian relation,
\be
	Y^{-m}_\ell = (-1)^m Y^{m *}_\ell,
\label{herm}
\ee
is automatically satisfied.

\subsection*{A.2 Normalisation and boundary conditions}
The $c_{\ell n}$ are normalization constants introduced to preserve 
the normalisation of the spherical Bessel function in  a finite
sphere. They are derived via the application of the orthogonality relations for 
spherical Bessel functions, for given boundary conditions (see HT).
The boundary conditions used for this analysis are Neumann boundary
conditions (i.e. that the first derivative of the potential field
normal to the boundary is zero). This implies that the velocity field
is zero on the boundary i.e. the boundary is undistorted. We therefore
need not include surface effects. The boundary condition gives us the
wavenumbers $k_{\ell n}$:
\be
 \frac{d}{dr}j_{\ell}\left(k_{\ell n}r\right)\vert_{r=r_{\rm max}} = 0,
\ee
and the normalization constants
\be
 	c_{\ell n} = \frac{k_{\ell n}}{|j_\ell(k_{\ell n}r_{\rm max})|
	\sqrt{\frac{r}{2}\left[\frac{1}{4}
+ k^{2}_{\ell n}r^{2}_{\rm max} - \left(\ell +
\frac{1}{2}\right)^{2}\right]}}.
\ee

\subsection*{A.3 Redshift Distortions}
As described in the main text, the transformation to redshift space can
be carried out by the identity $d^3s\, \rho(\s)=d^3r\, \rho(\r)$,
leaving redshifted coordinates in the weighting and spherical Bessel 
functions. HT then expand $w(s)j_{\ell}(k_{\ell n}s)$ in equation (\ref{26}) 
to first order in $s-r$,
\be
	w(s)j_{\ell}(k_{\ell n}s) \simeq w(r)j_{\ell}(k_{\ell n}r) +
u({\bf{r}})\frac{d}{dr}\left[w(r)j_{\ell}(k_{\ell n}r)\right],
\label{distort}
\ee
where $u(\r)$ is the radial component of the peculiar velocity field, 
$\vb(\r)-\vlg$.
Linear theory provides a relation between the peculiar velocity field
$\vb(\r)$ and the density field via the continuity equation
\be
 \nabla.\v(\r) = - H_{0}\Omega_{0}^{0.6}\delta_{\rho}(\r)
\ee
to first order.
In Appendix A of HT it is shown that the radial component of this field is 
\be
	u(\r)   = \Omega_{0}^{0.6} \sum_{\ell mn} c_{\ell n} k^{-1}_{\ell n}
	\delta_{\ell mn} j'_{\ell} (k_{\ell n} r) Y_{\ell m}
	\left( \theta,\phi \right) - \vlg \cdot \rhat,
\label{expand}
\ee
where $j'(z) =dj_\ell(z)/dz$ and $\delta_{\ell mn}$ is the transform
of the galaxy number over-density field $\delta(\r)$. Note that in 
this expression we have added
the contribution from the Local Group dipole. This was neglected in 
the analysis of HT since it only contributed significantly to the 
$\ell=1$ mode, which was not included in the analysis. We include it here 
for completeness.

Substituting equation (\ref{expand}) into equation (\ref{distort}) gives
\be
	w(s)j_{\ell}(k_{\ell n}s)  \simeq w(r)j_{\ell}(k_{\ell n}r) 
		+ \beta \sum_{\ell'm'n'}
	c_{\ell' \! n'}k_{\ell'\!n'}^{-1}\delta_{\ell'\!m'\!n'}
		j'_{\ell'}(k_{\ell'\!n'}r)
	\frac{d}{dr}\left[w(r)j_{\ell}(k_{\ell n}r)\right]Y_{\ell'\!m'} -
	(\vlg \cdot \rhat) \frac{d}{dr}\left[w(r)j_{\ell}(k_{\ell n}r)\right].
\label{34}
\ee

\subsection*{A.4 Transforming real space harmonic modes to redshift space}

From the definition of perturbations in the galaxy density field
we have
\be
\rho(\r) = \rho_{0}(\r)\left[1 + \delta(\r)\right] 
= \rho_{0}(\r)\left(1 +
\sum_{\ell mn}c_{\ell n}\delta_{\ell mn}j_{\ell}\left(k_{\ell n}r\right)
Y_{\ell m}\left(\theta,\phi\right)\right),
\label{35}
\ee
where $\rho_0(\r)$ is the observed mean density of galaxies in the 
survey.

Substituting equations (\ref{34}) and  (\ref{35}) into equation (\ref{26})
we find 
\be
\hat{\rho}_{\ell mn} = \left(\rho_{0}\right)_{\ell mn} +
\sum_{\ell'\!m'\!n'}\left(\Phi^{m
m'}_{\ell \ell'nn'} +
\beta V^{mm'}_{\ell \ell'nn'}\right)\delta_{\ell'\!n'}^{m'}.
\label{result}
\ee
The mean--field harmonics are defined as
\be
	\left(\rho_{0}\right)_{\ell mn} =
c_{ln} \int d^3\!r \,  \rho_{0}(\r) w(r) j_{\ell} \left(k_{\ell n}r \right)
Y^*_{\ell m}\left(\theta,\phi\right) - c_{ln}\int \!d^3\!r \,(\vlg \cdot \rhat)
\rho_{0}(\r){{\rm d}\over {\rm d}r}\left[w(r)j_{\ell}\left(k_{\ell n}r\right)
\right] Y^*_{\ell m}\left(\theta,\phi \right),
\ee
where we have incorporated the the contribution of the dipole term, 
since this is independent of $\delta(\r)$. In the rest of the analysis  
we assume a Local Group velocity of 622 km s$^{-1}$ towards $\ell=
277^\circ$, $b=30^\circ$.

The remaining matrices in equation (\ref{result}) account for the 
mixing of modes due to the angular and radial geometry of the survey:
\be
\Phi^{mm'}_{\ell \ell'nn'}  =
c_{\ell n}c_{\ell' n'}\int \!d^3 r \,\rho_{0}(\r)w(r)j_{\ell}
\left(k_{\ell n}r\right)
  Y^*_{\ell m}\left(\theta,\phi\right)j_{\ell'}\left(k_{\ell'n'}r\right)
Y_{\ell'm'}\left(\theta,\phi\right)
\ee
and the angular geometry and radial distortion:
\be
	V^{mm'}_{\ell \ell'nn'} = 
	\frac{c_{\ell n}c_{\ell'n'}}{k^{2}_{\ell' n'}}
	\int \! d^3r\rho_{0}(\r)\frac{d}{dr}
	\left[w(r)j_{\ell}\left(k_{\ell n}r\right)\right] 
	Y^*_{\ell m}\left(\theta,\phi\right)j'_{\ell'}
	\left(k_{\ell'n'}r\right)Y_{\ell'm'}\left(\theta,\phi\right).
\ee

Defining the observable data vector $D_{\ell m n} = (\hat{\rho}_{\ell mn} -
(\hat{\rho}_{0})_{\ell mn})/\overline{\rho}$, and $\bdelta$, the vector
of underlying density field values, equation (\ref{result}) can be 
written in a compact form as
\be
\D =(\bPhi + \beta \bV) \bdelta.
\ee
where we have adopted a matrix notation.

\subsection*{A.5 Separation into angular and radial terms}
The mean observed density field, $\rho_{0}(\r)$, may be separated into 
angular and radial components
\be
\rho_{0}(\r) = M\left(\theta,\phi\right)\rho_{0}(r),
\ee 
with $M(\theta,\phi)$ being the sky mask of the survey and
$\rho_{0}(r)$ containing the radial selection function. 
The `transition matrices' $\bPhi$ and $\bV$ now separate to become
\be
\Phi^{mm'}_{\ell \ell'nn'} =
W^{mm'}_{\ell \ell'}\Phi^{nn'}_{\ell \ell'}
\ee
\be
V^{mm'}_{\ell \ell'nn'} =
W^{mm{\prime}}_{\ell \ell'}V^{nn'}_{\ell \ell'}
\ee
where
\be
\Phi^{nn'}_{\ell \ell'} =
c_{\ell n}c_{\ell 'n'}\int^{r_{\rm max}}_0 \! r^{2}dr \, 
\rho_{0}(r)w(r)j_{\ell'}\left(k_{\ell'n'}r\right)
j_{\ell}\left(k_{\ell n}r\right),
\ee
\be
V^{\ell \ell'}_{nn'}  = 
\frac{c_{\ell n}c_{\ell 'n'}}{k^{2}_{\ell'n'}}\int^{r_{\rm max}}_0 \!
 r^{2}dr \, \rho_{0}(r)\frac{d}{dr}\left[w(r)j_{l}\left(k_{ln}r\right)\right]
 \frac{d}{dr}j_{\ell '}\left(k_{\ell 'n'}r\right)
\ee
and 
\be
W^{mm'}_{\ell \ell'} =
\int_{4 \pi}Y_{\ell'}^{m'}(\Omega) M(\Omega)Y^{*m}_{\ell}(\Omega)d\Omega.
\ee

\subsection*{A.6 ``Fingers of god'' and nonlinearities}

Throughout the paper we have also applied the correction (described by
HT) for the effect of the small scale non-linear peculiar
velocity field (see their Appendix B). In this procedure, it is assumed that 
the real-space position of a galaxy is perturbed slightly
by a peculiar velocity such that the radial redshift-space position becomes
\be 
 s(\r) \rightarrow s'(\r) = s(\r) + \epsilon (\r).
\ee
The quantity $\epsilon (\r)$ is a random variable drawn from a
Maxwellian distribution with mean zero and 
$\lgl \epsilon(\r) \epsilon(\r') \rgl
= (\sigma_v^2/3H_{0}^{2}) \delta_D(\r-\r')$ and where $\sigma_v$ is the 
3-d velocity 
dispersion. Errors in the measurement of redshift, $\sigma^2_z$
 can be incorporated into the formalism by adding the variances in quadrature;
$\sigma^2_{total}= (\sigma_v^2/3)+ \sigma^2_z$. Since this distortion is 
incoherent, the effect is to convolve each mode with a radial Gaussian
distribution. From this HT derived a scattering matrix $\Smat$ defined by
\be
S^{m m'}_{\ell \ell' n n'}  = 
\frac{c_{\ell n}c_{\ell'n'}\delta^{K}_{\ell \ell'}\delta^{K}_{mm'}}{V\pi}
\int \int
\frac{e^{-\frac{\left(r-y\right)^{2}}{\left(2\sigma\right)^{2}}}}{\sqrt{2\pi}\sigma}j_{\ell}\left(k_{\ell n}r\right)j_{\ell'}\left(k_{\ell' n'}y\right)
\,r dr \,y dy
\label{Smatrix}
\ee
which is to be convolved with the density field. $\delta^K$ is the
Kronecker delta function.  The amplitudes, $D_{\ell mn}$,
corrected for a small--scale incoherent velocity field are now given by
\be
D_{\ell mn} = S^{m m''}_{\ell \ell'' nn''}
W^{mm'}_{\ell''\ell'}\left(\Phi^{n''n'}_{\ell''\ell'}
+ \beta
V^{n''n'}_{\ell''\ell'}\right)\delta_{\ell'\!m'\!n'},
\ee
or in matrix notation,
\be
	\D= \Smat \W (\bPhi + \beta \bV) \bdelta.
\ee

\subsection*{A.7 Correlations}

HT assumed that the sky mask was azimuthally symmetric,
making the mask mixing matrices $\W$ real. In the current analysis, we
drop this assumption as the sky mask for the PSCz survey is not
symmetric (see Figure~\ref{psczsky}).  This complicates the analysis,
since the noise matrix is no longer real. We can avoid this problem by 
constructing the covariance matrix for the real and imaginary parts 
of $\D$. Since these are also Gaussian random, there are now four terms to
the covariance matrix $\C$,
\ba
\C_{\Real\Real} &=& \langle \Real \D \, \Real \D^t\rangle,\\
\C_{\Real\Imag} &=& \langle \Real \D \, \Imag \D^t\rangle, \\
\C_{\Imag\Real} &=& \langle \Imag \D \, \Real \D^t\rangle,\\
\C_{\Imag\Imag} &=& \langle \Imag \D \, \Imag \D^t\rangle, 
\ea
where $\Real$ signifies the real part and $\Imag$ the imaginary part. These 
terms are given by the following summations:
\be
	C_{\Real\Real}   =  \frac{1}{2} \sum_\alpha
	\left[\Real(\Phi^\alpha_\mu +
	\beta V^\alpha_\mu)\Real(\Phi^\alpha_\nu + \beta
	V^\alpha_\nu)  +    \Imag(\Phi^\alpha_\mu + \beta
	V^\alpha_\mu) \Imag(\Phi^\alpha_\nu + \beta
	V^\alpha_\nu)\right]P(k_\alpha) +  N_{\mu\nu},
\label{60}
\ee
\be
	C_{\Imag\Imag}  =  C_{\Real\Real},
\ee
\be	
	C_{\Real\Imag}  =  \frac{1}{2} \sum_{\alpha}\left[\Real(\Phi^\alpha_\mu +
	\beta V^\alpha_\mu)\Imag(\Phi^\alpha_\nu + \beta
	V^\alpha_\nu) - \Imag(\Phi^\alpha_\mu + \beta
	V^\alpha_\mu)\Real(\Phi^\alpha_\nu + \beta
	V^\alpha_\nu)\right]P(k_\alpha) + N_{\mu\nu}, 
\ee
\be
	C_{\Imag\Real}  =  -C_{\Real\Imag},
\label{63}
\ee
where $\mu=(\ell mn)$, $\nu=(\ell' m' n')$ and $\alpha=(\ell''  m'' n'')$.
The noise matrix, $\N$, is given by the equation 
\be
	N_{\mu \nu} = c_{\ell n}c_{\ell' n'}\int
	\rho_{0}(r)w^{2}(r)j_{\ell}\left(k_{\ell n}r\right)
	j_{\ell'}\left(k_{\ell'n'}r\right)r^{2}dr
 	\int{\cal{P}_{\mu}}
	\left[Y_{\ell m}(\Omega)\right] M(\Omega) {\cal{P}_{\nu}}
	\left[Y^{\star}_{\ell' m'}(\Omega)\right]d\Omega
\label{noise}
\ee
as in HT, where the ${\cal{P_{\mu,\nu}}}$ represent the real
or imaginary parts of the spherical harmonics according to whether we
are using the real or imaginary parts of $D_{\mu,\nu}$. In Appendix B
we discuss in more detail the construction of the mask and its decomposition
into correlations between real and imaginary parts of the field $\D$.

Now we have the covariance matrix, a likelihood functional is
constructed and we use an assumed real-space power spectrum $P(k)$ in
equations (\ref{60}) to (\ref{63}). We maximize the likelihood for the
value of $\beta$ and the amplitude of the power spectrum, as described
in the main text. The amplitude of the power spectrum is defined in
terms of $\Delta_{0.1}$ where $\Delta^{2}(k) =
k^{3}P(k)/2\pi^{2}$. The likelihood function (unnormalised) we use is
\ba
{\cal{L}}[\D\vert \beta, P(k_{ln})] &=& \left[\det(\C_{\Real\Real}) 
	\det(\C_{\Real\Imag}) \det(\C_{\Imag\Real})
	\det(\C_{\Imag\Imag})\right]^{-\frac{1}{2}} \nn
	&\times&
	\exp\left( 
	-\frac{1}{2}\Real \D^t \C^{-1}_{\Real\Real} \Real \D -
	\frac{1}{2}\Imag \D^t \C^{-1}_{\Imag\Real} \Real \D\right) \nn
	&\times& 
	\exp\left(-\frac{1}{2}\Real \D^t \C^{-1}_{\Real\Imag}\Imag \D-
	\frac{1}{2}\Imag \D^t \C^{-1}_{\Imag\Imag} \Imag \D\right).
\ea

\section*{Appendix B: Construction of the angular window matrix}

In this appendix we describe in detail the formulation and calculation of the 
angular window matrix. Since we drop the assumption that the angular mask
is axisymmetric, calculation of the data covariance matrices is somewhat
more complicated than that found in HT (see Appendix A.7).
In this Appendix we define the window transition matrix, $\W$, and 
describe how we can calculate it efficiently with the use of Clebsch--Gordan,
or Wigner 3j coefficients. We then show that formally the data covariance
matrix is imaginary, due to the non-axisymmetric mask. We expand the 
window matrix and show how the data covariance matrix can be made positive
definite by taking the covariances of the real and imaginary parts of the
data vector, $\D$. Finally, we present the explicit expressions used
to calculate the full covariance matrix, including the noise terms.

\subsection*{B.1 Definitions and conventions}
  
        An angular mask function is defined as
\be
        M(\Omega) = \left\{ \begin{array}{ll}
                        1 & \mbox{if observable area of sky,} \\
                        0 & \mbox{otherwise}
                            \end{array}
                        \right.
\ee
From this we can define the transition matrix
\be
        W_{\ell \ell'}^{m m'} = \int_{4 \pi} 
                        Y_{\ell'}^{m'}(\Omega) M(\Omega) Y_{\ell}^{m*}(\Omega)
                        \,  d\Omega  ,
\label{eq1}
\ee
In this Section  we shall abbreviate our notation by dropping 
$\ell$, $m$ subscripts and $\int\, d\Omega$ 
except where needed for clarity. Note our convention for dashes and
complex conjugates in the definition of $\W$.
The $\W$ matrix operates on the density field by mixing modes;
\be
	D'_{\ell m n} = \sum_{\ell'm'}  W_{ll'}^{mm'} D_{\ell' m' n}.
\ee
Rather than directly performing the integration in equation 
(\ref{eq1}), which is a prohibitively slow process for large $\ell$, 
it is faster to decompose the calculation into a two-step process,
as outlined by \scite{SHLLB92}. First one calculates the single 
transform of the mask function 
\be
	        M_{\ell m} = \int_{4 \pi} M(\Omega) Y_{\ell}^{m*}(\Omega)
                        \,  d\Omega  ,\\
\ee
which can then be use to calculate the full transition matrix by
the transformation ($Y' \equiv Y_{\ell'}^{m'}$ etc.)
\be
        W_{\ell\ell'}^{mm'} = \sum_{\ell''m''} M_{\ell''}^{m''}
                        \int Y' Y'' Y^{*}.
\ee 

\subsection*{B.2 Clebsch--Gordan \& Wigner 3j coefficients}

The angular integral, $\int Y' Y'' Y^{*}$
can be calculated analytically by the Clebsch--Gordan or Wigner $3j$ 
matrices \cite{Edmonds}. The  Wigner $3j$ symbols are related to the  
Clebsch--Gordan coefficients by
\be
        \left(
        \begin{array}{ccc}      \ell & \ell' & \ell'' \nn
                                  m &   m'  &   m''
        \end{array}
         \right) = 
	(-1)^{\ell-\ell'-m''}(2 \ell''+1)^{-1/2}
	 \lgl \ell m\ell'm'|\ell\ell'\ell''-m'' \rgl
\ee
Here we choose to express relations in terms of the $3j$ matrices. Hence 
\ba
\int \! Y Y' Y'' =   
 \sqrt{ \frac{(2\ell+1)(2\ell'+1)(2\ell''+1)}{4 \pi} }
        \left(
        \begin{array}{ccc}      \ell & \ell' & \ell'' \nn
                                  0 &   0  &   0
        \end{array}
        \right)
        \left(
        \begin{array}{ccc}      \ell & \ell' & \ell'' \nn
                                  m &   m'  &   m''
        \end{array}
         \right)
\ea
with the triangle conditions $m+m'+m''=0$ and
$|\ell-\ell'|\leq \ell'' \leq (\ell+\ell')$,
plus permutations. Integrals containing complex conjugates of $Y$'s can be
derived using the 
hermicity of the spherical harmonics (equation (\ref{herm})).

 The  Wigner $3j$ symbols are defined by \scite{LL77}
\ba
        \left(
        \begin{array}{ccc}      \ell & \ell' & \ell'' \nn
                                  m &   m'  &   m''
        \end{array}
         \right) = &\delta^K_{m_1+m_2,m}& (-1)^{\ell-\ell'-m''} 
	\sqrt{\frac{(\ell +\ell' - \ell'')! 
		(\ell -\ell' +\ell'')! (- \ell + \ell'+\ell'')!}{
		(\ell+\ell'+\ell''+1)!}} \nn
	& &
	[(\ell+m)!(\ell-m)!(\ell'+m')!(\ell'-m')!
	(\ell''+m'')!(\ell''+m'')!]^{1/2} 
	\nn & &
	\sum_{z} (-1)^z [ z!  (\ell +\ell' - \ell''-z)! 
		(\ell-m-z)! (\ell'+m'-z)!
	\nn & &	
	 (\ell''-\ell'+m+z)!(\ell''-\ell'-m'+z)!]^{-1}
\ea
where the summation over $z$ is zero when any factorial terms in the 
summation are negative. When $m=m'=m''=0$ then 
\ba
	 \left(
        \begin{array}{ccc}      \ell & \ell' & \ell'' \nn
                                  0 &    0   &   0
        \end{array}
         \right) &=& (-1)^{L/2} 
		\sqrt{\frac{(L-2\ell)!(L-2\ell')!(L-2\ell'')!}{(L+1)!}} 
			\frac{(L/2)!}{(L/2-\ell)!(L/2-\ell')!(L/2-\ell'')!}, 
		\hspace{0.6in}\ell \hspace{0.5cm}{\rm even}, \nn
		&=& 0, \hspace{4.in}  \ell \hspace{0.5cm} {\rm odd}.
\ea
where $L=\ell+\ell'+\ell''$.
Practically it is easier to calculate these terms from 
$ \left(
        \begin{array}{ccc}      \ell & \ell &   0 \nn
                                  0 &    0   &   0
        \end{array}
         \right)$ using the 
recursive relations
\be
 	 \left(
        \begin{array}{ccc}      \ell & \ell' & \ell'' \nn
                                  0 &    0   &   0
        \end{array}
         \right) = 
	\sqrt{\frac{(L-2\ell'-1)(L-2\ell''+2)}{(L-2\ell')(L-2\ell''+1)}}
	\left(
        \begin{array}{ccc}      \ell & \ell'+1 & \ell''-1 \nn
                                  0 &     0    &   0
        \end{array}
         \right).
\ee
With these relations, the computation time for the mask mixing 
matrix is reduced from days to a few minutes for $\ell$ up to $100$.

\subsection*{B.3 Covariance Matrix Terms}

	Armed with the basic definitions of the window function
transformation, and methods to generate it efficiently, we now begin to 
enumerate the various terms that will appear in the data covariance matrix. 
As stated above the problem of constructing a data covariance matrix is 
complicated by the combined presence of noise and a non--axisymmetric mask,
leading to an imaginary covariance matrix. This can be dealt with
in linear theory by decomposing the harmonic modes into real and 
imaginary parts and calculating their covariances. We begin with a 
formal derivation of the data covariance matrix.

The covariance matrix of the observed modes can be written 
in matrix notation as (c.f. equation 
(\ref{modemode}))
\be
\C =\lgl \D \D^\dag \rgl = 	
\Smat  \W (\bPhi+\beta \bV) \bP  (\bPhi+\beta \bV)^t \W^\dag \Smat^t +  
	\R \W,
\label{fullcov}
\ee 
where $\bP={\rm diag}[P(k_{\ell n})]$ and
the shot noise term is given by $\N=\R \W$, which we have 
decomposed into radial and angular parts (c.f. equation (\ref{noise})), 
\be
	\R=c_{\ell n}c_{\ell'n'}\int
	\rho_{0}(r)w^{2}(r)j_{\ell}\left(k_{\ell n}r\right)
	j_{\ell'}\left(k_{\ell'n'}r\right)r^{2}dr
\ee
and
\be
	\W = \int Y' M Y^*.
\ee
Note that here we use the full window transition matrix.

 Expanding
$\W$ into real and imaginary parts, $\W = \Real \W + i \Imag \W$, yields 
\ba
	\C &=& 
\Smat\,\Real \W(\bPhi+\beta \bV)\bP(\bPhi+\beta \bV)^t\,\Real \W^t\Smat^t +
\Smat\,\Imag \W(\bPhi+\beta \bV)\bP(\bPhi+\beta \bV)^t\,\Imag \W^t\Smat^t +
	\R \, \Real \W \nn &+& \!\!i
\Smat\,\Imag \W(\bPhi+\beta \bV)\bP(\bPhi+\beta \bV)^t\,\Real \W^t\Smat^t -i
\Smat\,\Real \W(\bPhi+\beta \bV)\bP(\bPhi+\beta \bV)^t\,\Imag \W^t\Smat^t +
  	i \R \,\Imag \W
\label{eq:imagcov}
\ea
where the real and imaginary parts of $\W$ can also be written
\be
	\Real \W = \Sym \W,
\ee
and
\be
	\Imag \W =\frac{1}{i} \Asym \W,
\ee
where the operators $\Sym$ and $\Asym$ symmetrize and anti-symmetrize
the matrices.

\subsection*{B.4 Expansion of the window matrix}

Although imaginary terms exist in the data covariance matrix, we 
can use the property of a Gaussian field that the real and imaginary
parts will be independent and each is independently Gaussian distributed 
to construct total the probability distribution function. We later 
show that the covariance matrix can be further expanded into 
independent terms that will allow us to define a total covariance 
matrix which is both real and symmetric.

 Expanding the $Y$'s into real and imaginary parts we find that 
\be
	\W = 	(\Real \Real \W + \Imag \Imag \W) + 
		i (\Real \Imag \W - \Imag \Real \W),
\ee
where we introduce the notation $\Real \Real \W = \int \Real Y M \Real Y'$,
 $\Imag \Imag \W =\int \Imag Y M  \Imag Y'$, 
$\Real \Imag \W = \int\Real Y M \Imag Y'$
and  $\Imag \Real \W = \int\Imag Y M \Real Y'$. The real and imaginary 
parts of the spherical harmonics can be found from
\be
	\Real Y = \frac{1}{2}(Y+Y^*)
\ee
and
\be
	\Imag Y = \frac{1}{2i}(Y-Y^*)
\ee
The terms $\Real \Real \W$ can then be calculated from the full window
matrix by
\ba
	\Real \Real \W &=& \frac{1}{4} \left( \int YMY'+\int Y^*MY' 
		+\int Y M Y'^{*}+ \int Y^* M Y'^{*} \right), \nn
	\Imag \Imag \W &=& \frac{1}{4} \left( \int Y^*MY'+\int YMY'^{*} 
		-\int Y M Y'- \int Y^* M Y'^{*}\right), \nn
	\Real \Imag \W &=& \frac{1}{4i} \left( \int YMY'-\int YMY'^{*} 
		+\int Y^* M Y'- \int Y^* M Y'^{*}\right), 
\ea
where the various permutations can be calculated from the window function 
matrix using the Hermicity relations.
These are the terms that 
will appear in the noise matrix, equation (\ref{noise}). Note the ordering 
of dashed terms as these terms are not in general symmetric in their indices. 
Clearly then $\Real \W = \Real \Real \W + \Imag \Imag \W$ and 
$\Imag \W = \Real \Imag \W - \Imag \Real \W$.

\subsection*{B.3 Transpose relations}

Here we list without proof the transpose relations for the three
window matrices.
\ba
	\Real \Real \W &=& (\Real \Real \W)^T , \nn
	\Imag \Imag \W &=& (\Imag \Imag \W)^T, \nn
	\Imag \Real \W &=& (\Real \Imag \W)^T.
\ea
Given the symmetry relation $\Imag \Real \W = (\Real \Imag \W)^T$,
 we only need explicitly to calculate the three window matrices 
$\Real \Real \W$, $\Imag \Imag \W$ and $\Imag \Real \W$.

\subsection*{B.4 Parity Relations}

Here we list without proof the parity relations ($m \rightarrow -m$)
for the three basic window matrices.
\ba
	\Real \Real W_{m,-m'} &=& (-1)^{m'} \Real \Real W_{m m'} , \nn
	\Real \Real W_{-m,m'} &=& (-1)^{m} \Real \Real W_{m m'} , \nn
	\Real \Real W_{-m,-m'} &=& (-1)^{m'+m} \Real \Real W_{m m'} , 
\ea
\ba	
	\Imag \Imag W_{m,-m'} &=& (-1)^{m'+1} \Imag \Imag W_{m m'} , \nn
	\Imag \Imag W_{-m,m'} &=& (-1)^{m+1} \Imag \Imag W_{m m'} , \nn
	\Imag \Imag W_{-m,-m'} &=& (-1)^{m+m'} \Imag \Imag W_{m m'} , 
\ea
\ba
	\Imag \Real W_{m,-m'} &=& (-1)^{m'} \Imag \Real W_{m m'} , \nn
	\Imag \Real W_{-m,m'} &=& (-1)^{m+1} \Imag \Real W_{m m'} , \nn
	\Imag \Real W_{-m,-m'} &=& (-1)^{m+m'+1} \Imag \Real W_{m m'} . 
\ea

\subsection*{B.5 Expansion of covariance matrix}

We can re-arrange the covariance function by reducing
the data vector into real and imaginary parts; $\D = \Real \D +i \Imag \D$.
This is convenient because our hypothesis is that $\D$ is a Gaussian
random variable and so the terms $\Real \D$ and $\Imag \D$ will also 
be Gaussian random variables. In this case we need terms like
$\lgl \Real \D \, \Real \D'\rgl$.
This leads us to
\be
	\C = (\lgl \Real \D \, \Real \D'^{t} \rgl + 
	\lgl \Imag \D \, \Imag \D'^{t} \rgl)
 	+ i(\lgl \Imag \D \, \Real \D'^{t} \rgl - 
	\lgl \Real \D \, \Imag \D'^{t} \rgl).
\label{eq:exp1}
\ee
 We can further expand terms 
by expanding the real and imaginary terms of the $\W$ matrix and 
the underlying field, $\delta$. Hence 
$\Real D = \Real \W \, \Real \delta -\Imag \W \, \Imag \delta$ and 
$\Imag D = \Imag \W \, \Real \delta + \Real \W \, \Imag \delta$.
Under the assumption of statistical independence between the real and 
imaginary $\delta$'s we find
\ba
	\lgl \Real \D \, \Real \D^t \rgl &=& 
	\frac{1}{2} 
	\Smat\,\Real\W(\bPhi+\beta\bV)\bP(\bPhi+\beta\bV)^t\,\Real\W^t\Smat^t+
	\Smat\,\Imag\W(\bPhi+\beta\bV)\bP(\bPhi+\beta\bV)^t\,\Imag\W^t\Smat^t+ 
	\R \, \Real \Real \W , \nn
	\lgl \Imag \D \, \Imag \D^t \rgl &=& 
	\frac{1}{2}
	\Smat\,\Real\W(\bPhi+\beta\bV)\bP(\bPhi+\beta\bV)^t\,\Real\W^t\Smat^t+
	\Smat\,\Imag\W(\bPhi+\beta\bV)\bP(\bPhi+\beta\bV)^t\,\Imag\W^t\Smat^t+ 
	\R \, \Imag \Imag \W , \nn
	\lgl \Real \D \, \Imag \D^t \rgl &=& 
	\frac{1}{2}	
	\Smat\,\Real\W(\bPhi+\beta\bV)\bP(\bPhi+\beta\bV)^t\,\Imag\W^t\Smat^t-
	\Smat\,\Imag\W(\bPhi+\beta\bV)\bP(\bPhi+\beta\bV)^t\,\Real\W^t\Smat^t- 
	\R \, \Real \Imag \W , \nn
	\lgl \Imag \D \, \Real \D^t \rgl &=& 
	\frac{1}{2}	
	\Smat\,\Imag\W(\bPhi+\beta\bV)\bP(\bPhi+\beta\bV)^t\,\Real\W^t\Smat^t-
	\Smat\,\Real\W(\bPhi+\beta\bV)\bP(\bPhi+\beta\bV)^t\,\Imag\W^t\Smat^t- 
	\R \, \Imag \Real \W . \nn 
\ea 
Substituting these expressions into equation (\ref{eq:exp1}) we reproduce
equation (\ref{eq:imagcov}). In addition we have shown that each of these
terms are independent and Gaussian distributed. Therefore we
can construct a real covariance matrix;
\be
	\C =    
        \left(
        \begin{array}{cc}      \C_{\Real \Real} & \C_{\Real \Imag} \nn
                               \C_{\Imag \Real} & \C_{\Imag \Imag}  
        \end{array}
         \right)
\ee
From the symmetry relations given earlier we see that this is 
a real, symmetric matrix.

\end{document}